\documentclass{elsart}

\usepackage{graphicx}

\usepackage{harvard}
\usepackage{amssymb}

\def\bibcode#1{(\texttt{#1})}

\newcommand{\vr}{\mbox{\boldmath $r$}}

\begin{document}

\begin{frontmatter}
\title{Contribution of the Large Magellanic Cloud to the Galactic Warp}
\author{Toshio Tsuchiya\thanksref{email}}
\address{Astronomisches Rechen-Institute, M\"{o}nchhofstra{\ss}e 12--14,
  D69120 Heidelberg, Germany}
\thanks[email]{E-mail: tsuchiya@ari.uni-heidelberg.de}

\begin{abstract}
  Multi-scale interaction between the LMC, the Galactic halo, and the disk
  is examined with N-body simulations, and precise amplitudes of the
  Galactic warp excitation are obtained. The Galactic models are
  constructed most realistically to satisfy available observational
  constraints on the local circular velocity, the mass, surface density
  and thickness of the disk, the mass and size of the bulge, the local
  density of the halo matter at the solar radius, and the mass and orbit
  of the LMC\@.  The mass of the halo within $R=50$~kpc is set to about
  $5 \times 10^{11}M_\odot$. Since the observational estimate of the mass
  distributed in outer region has large ambiguity, two extreme cases are
  examined; $M(<170\mathrm{kpc}) = 2.1$ and $0.9\times10^{12}M_\odot$.
  LMC is orbiting in a ellipse with apocentric radii of 100~kpc, thus
  the main difference between our two models is the mass density in the
  satellite orbiting region, so that our study can clarify the role of
  the halo on excitation of the warp.
  
  By using hybrid algorithm (SCF-TREE) I have succeeded to follow the
  evolution with millions of particles. The orbiting satellite excites
  density enhancement as a wake, and the wake exerts a tidal force on
  the disk. Because of the additional torque from the wakes in the halo,
  the amplitudes of the induced warps are much larger than the classical
  estimate by \citeasnoun{hunt1969}, who considered only the direct
  torque from the LMC\@. The obtained amplitudes of $m=0$, 1, 2 warps in
  the larger halo model show very good agreement with the observed
  amplitude in the Milky Way. This result revives the LMC as a possible
  candidate of the origin of the Galactic warp. Our smaller halo model,
  however, yield only weak warps in all the harmonic modes. Therefore
  the halo still has significant influence on excitation of warp even
  in the interaction scenario for excitation of warps.
\end{abstract}

\end{frontmatter}

\newpage
\section{Introduction}
\label{sec:Introduction}

The origin of galactic disk warping, which is typically vertical bending of
disks in a integral shape, is a long-standing problem. Even after 40
year efforts to explain the phenomena since the discovery of the warp
in the Milky Way \cite{burke1957,kerr1957}, there is still no generally
acceptable theory. A general review of the problem can be found in
\citeasnoun{binn1992}.

There are several observational facts that any convincing theory must
explain. The first point is the longevity of the warps. Observations of neutral
hydrogen layers \cite{bosm1991} and stellar disks
\cite{sanc1990,resh1998} have revealed that about a half of spiral
galaxies exhibit warps. Such high frequency of warps indicate that the
warps are long-lived with about the same age as galaxies, otherwise some
mechanism keeps exciting the warps. In both cases the warp excitation
mechanism must be ubiquitous. The second point is the warp amplitude.
For instance the Milky Way has a prominent warp and its amplitude is
2~kpc at Galactocentric distances $R\sim14$~kpc\cite{hend1982}, and 4~kpc
at $R\sim 20$~kpc \cite{dipl1991}. The torque that can bend the disk in
such a large amplitudes is so strong, which make it difficult to find
the source of the torque \cite{hunt1969}. The third point is
alignment of the warp nodes. Most of spiral galaxies including our own
have straight line of nodes within the Holmberg radius, and outward the
lines of nodes make leading spirals \cite{brig1990}. This fact presents
a contrast to the spiral arms in the grand-design spiral.

Simple analysis \cite{binn1992} shows that the simplest interpretation
of the warp as a vertical oscillation of matters without considering
selfgravity, is not working, because the line of nodes wind up in a
time scale of 1 or 2~Gyr. Therefore self-consistent treatment is
necessary. There are several hopeful theories, all of which have,
however, some shortcomings.

The first step to understand long-lived warps is the normal mode
analysis in steady state galaxies. \citeasnoun{hunt1969} showed that
there are many bending normal modes in their thin-disk models, but
unless density of the disks are cut-off sharply at their edges the
frequency spectrum is at least partly continuous. In such disks any
realistic perturbation would excite propagating wave packets, thus the
warping wave is dispersed. A long-lived warp can correspond only to a
discrete mode. If the disk is truncated at some radius, there is only
one discrete normal mode, which is called a {\em modified tilt
  mode} \cite{spar1988}. \citeasnoun{spar1988} showed that if disk is
embedded in a flattened halo there is at most one discrete mode. Shapes of
the mode show good agreement with observed warps, and even though the
disk is not in the right shape of the modified tilt mode initially, it
dynamically evolves to fit the mode \cite{hofn1994}.

The normal mode theory works very well as long as the halo is treated as
a rigid body. However, once the self-consistent response of halo is
included several problems emerge. For example, dynamical friction acting
on the disk from the halo dissipates the warping motion
\cite{nels1995,dubi1995}, and also the line of nodes often winds up
tightly \cite{binn1998a,idet2000}. To find the normal modes in fully
self-consistent composite systems of a disk and a halo is extremely
intractable.

One possible mechanism to keep the misalignment between a disk and a
halo symmetric plane is continuous addition of matter with misaligned
angular momentum to the halo. Inclined cosmic infall
\cite{ostriker-e1989,jian1999} might support the mechanism.
\citeasnoun{jian1999} have shown that when infall reorientates the outer
part of a galactic halo by several degrees per Gyr, a realistic warp is
excited. Though the result seems appealing, their model of the infall is
still too idealistic, and numerical studies are relatively short ($\sim$
1~Gyr) to explain long-lived warps. Refined studies are necessary.

Another mechanism, which we are examining in this paper, is interaction
with satellites. This mechanism is intuitively plausible because the
most frequent $m=1$ warp is easily associated with a tidal field from a
satellite. Furthermore increasing observational evidence suggests that
most of the galaxies have satellite dwarfs. In fact, recently a faint
dwarf companion has been discovered around NGC 5907, which was
considered as a typical example of a warping galaxy without interacting
companions \cite{shan1998}. Also statistics show positive correlation
between warping and existence of interacting companions \cite{resh1998}

The only but severe problem of the interaction scenario is that the
tidal forces from satellites are usually too small. For example,
\citeasnoun{hunt1969} showed that the direct tidal force from the Large
Magellanic Cloud (LMC) of $10^{10}M_\odot$ at 50~kpc can excite a warp
with an amplitude at largest 100~pc at 16~kpc, while the real Galactic
warp is about 3~kpc.

Against the problem, a remedy was proposed by \citeasnoun{wein1998c}
(hereafter W1998).  Main point of his idea is to incorporate the halo
response to the orbiting motion of a satellite into the source of the
tidal forces. As the satellite moves, rotating density waves which have
resonant frequencies to the satellite rotation are excited. In a nearly
isothermal halo the most dominant wake appears at the position of 2:1
resonance, which is at about half distance to the satellite orbit. And
the mass of the wake is about the same as the satellite, thus its tidal
force is much larger than those from the satellite.  W1998 showed with
linear analysis for infinitely thin disks, that the tidal forces from
the Large Magellanic Cloud (LMC) and its wake are large enough to excite
the observable warp.

In order to prove this idea, fully self-consistent numerical simulations
of three-dimensional disks are necessary. This is, in fact, very
difficult and expensive, because we have to treat the small vertical motions
of a very thin disk embedded in a very massive halo, which is extended well
beyond the LMC orbit. W1998 estimated that 1 million
particles are necessary to reduce numerical noise in order to
distinguish meaningful dynamics of the disk.

This requirement is here accomplished by using a hybrid algorithm,
TREE-SCF \cite{vine1998}, which particularly suits our problem of warping
disk dynamics. Furthermore, we make the models as close to the real
Milky Way -- LMC system as possible, so that we can make precise
evaluation of the warp amplitudes.

In section~\ref{sec:Galaxy_Models} we first describe by a brief summary the
method to construct the Galaxy models using \citeasnoun{kuij1995},
then give some parameters of the models which we use in this study. The
precise input parameters in the \citeasnoun{kuij1995} codes are listed
in the appendix. Section~\ref{sec:Numerical_Methods} explains algorithms
and parameters of the TREE-SCF code. The results of the numerical
simulations are given in
Section~\ref{sec:Simulations}. Section~\ref{sec:Conclusion} is devoted
to conclusion and discussions.

\section{Galaxy Models}
\label{sec:Galaxy_Models}

The aim of this study is to examine the influence of the LMC on
excitation of the Galactic warp. Therefore we must first construct
equilibrium Galaxy models that have no warp, and then put a satellite in
the LMC orbit.

To construct equilibrium models for a multi-component system like the
Milky Way is actually a hard task. Many galactic models are constructed
with the assumptions that the halo and bulge are static backgrounds, or
assuming that velocity distribution is Gaussian and solving only the
Jeans equations. The deviation of these approximations from a true
equilibrium causes initial transient behavior, in other words relaxation
into the true equilibrium. That behavior changes the initial distribution
especially in the disk. In the disk warping problem it is necessary to
avoid any disturbance on the disk other than that from the satellite.

A suitable initial model is given by \citeasnoun{kuij1995} (hereafter
KD1995), which is nearly an exact solution of the collisionless
Boltzmann and Poisson's equations. In their method, the configurations
of a halo, bulge, and disk are given by means of distribution functions
which depend only on integrals of motion. By the Jeans theorem such
distribution functions are already steady state solutions of the
collisionless Boltzmann equation. Densities can be calculated from the
distribution functions, which depend on the positions only through the
potential function.  Therefore the main task is to determine the
functional form of the potential by solving Poisson's equation
numerically.

The halo distribution function takes the form
\begin{equation}
  \label{eq:haloDF}
  f_{\mathrm{halo}}(E,L_z^2) = \left\{
    \begin{array}{ll}
      \left[ (AL_z^2+B)\exp (-E/\sigma_0^2) + C\right]
      \left[ \exp (-E/\sigma_0^2) -1\right]  \quad&
      \mathrm{if} E<0, \\
      0 & \mathrm{otherwise}, \\
    \end{array}
    \right.
\end{equation}
where $L_z$ is the specific angular momentum about the axis of symmetry,
and $E$ is the relative energy, which is defined so that $E=0$ at the
edge of the distribution (where $\rho=0$)\cite{binn1987}. This
distribution function has five free parameters: the potential at the
center $\Psi_0$, which appears implicitly through the definition of the
relative energy, the velocity scale $\sigma_0$, and three factors $A$,
$B$, and $C$. This distribution is based on Evans's model, which has
axisymmetric logarithmic potential, but is truncated at a finite radius
by lowering the distribution function in the same manner as the lowered
isothermal models. The factors $A$ and $B$ control the system flattening
($q$) and the core radius ($R_c$), respectively, and all the three factors
are scaled by the density scale ($\rho_1$).

The bulge distribution function is the same as a King model
\cite{binn1987} and has the form
\begin{equation}
  f_{\mathrm{bulge}}(E)=\left\{ 
    \begin{array}{ll}
      \rho_\mathrm{b}(2\pi\sigma_\mathrm{b}^2)^{-3/2}
      \exp[(\Psi_0-\Psi_\mathrm{c})/\sigma_\mathrm{b}^2] &
      \{\exp[-(E-\Psi_\mathrm{c})/\sigma_\mathrm{b}^2]-1\} \\
      & \mbox{if } E<\Psi_\mathrm{c}, \\
      0 & \mbox{otherwise}.
    \end{array} \right.
\end{equation}
This distribution has 3 parameters. $\Psi_\mathrm{c}$ is the cutoff
potential, $\rho_\mathrm{b}$ the central bulge density, and
$\sigma_\mathrm{b}$ the bulge velocity dispersion.

For these two distributions, the densities are given by analytic
functions of $R$ and $\Psi$ \cite{kuij1995}.

The disk distribution must depend on 3 integrals of motion, in order to
keep the triaxial velocity ellipsoid in the disk \cite{dehn1998c}.
The third integral is, however, not known as any analytical function, so
that distribution function cannot be written as an analytical function.
For the sake of convenience, an approximated distribution function is
employed which depends on the energy in the vertical oscillations,
$E_z\equiv \Psi(R,z)-\Psi(R,0)+\frac{1}{2}v_z^2$, and that in planer
motion, $E_\mathrm{p}\equiv E-E_z$, as well as $L_z$. $E_z$ is quite
well conserved for stars in nearly circular orbits. The form is

\begin{equation}
  \label{eq:diskDF}
  f_{\mathrm{disk}}(E_\mathrm{p},L_z,E_z) =
  \frac{\Omega(R_\mathrm{c})}{(2\pi^3)^{1/2}\kappa(R_\mathrm{c})}
  \frac{\tilde{\rho_\mathrm{d}}(R_\mathrm{c})}
  {\tilde{\sigma_R}^2(R_\mathrm{c})\tilde{\sigma}_z(R_\mathrm{c})}
  \exp\left[ -\frac{E_p-E_\mathrm{c}(R_\mathrm{c})}
  {\tilde{\sigma}_R^2(R_\mathrm{c})} - 
  \frac{E_z}{\tilde{\sigma}_z^2(R_\mathrm{c})} \right]
\end{equation}
where $R_\mathrm{c}$ and $E_\mathrm{c}$ are the radius and energy of a
circular orbit with angular momentum $L_z$, and $\Omega$ and $\kappa$
are the circular and epicyclic frequencies at radius
$R_\mathrm{c}$. The `tilde' functions $\tilde{\rho_\mathrm{d}}$,
$\tilde{\sigma}_R$, and $\tilde{\sigma}_z$ are free functions. Therefore
the density and the radial velocity dispersion are conveniently selected as
\begin{equation}
  \label{eq:disk_density}
  \rho_\mathrm{disk}(R,z) = \frac{M_\mathrm{d}}{8\pi
    R_\mathrm{d}^2z_\mathrm{d}}\mathrm{e}^{-R/R_\mathrm{d}}
    \mathrm{erfc} \left(
    \frac{r-R_\mathrm{out}}{\sqrt{2}\delta R_\mathrm{out}} \right)
  \exp\left[ -4.6187\frac{\Psi_z(R,z)}{\Psi_z(R,3z_\mathrm{d})}
    \right],
\end{equation}
and
\begin{equation}
  \label{eq:disk_rad_velocity_dispersion}
  \tilde{\sigma}_R^2 = \sigma_R^2(0)\exp(-R/R_\mathrm{d}).
\end{equation}
$\tilde{\rho_\mathrm{d}}$ and $\tilde{\sigma}_z$ are iteratively
adjusted so that the densities on the mid-plane and at height
$z=3z_\mathrm{d}$ will agree with those given by
eq.~(\ref{eq:disk_density}).  The distribution of the disk has 6 free
parameters: the disk mass, $M_\mathrm{d}$, the radial scale length
$R_\mathrm{d}$, the vertical scale height $z_\mathrm{d}$, the disk
truncation radius $R_\mathrm{out}$, the truncation width $\delta
R_\mathrm{out}$, and the central velocity dispersion of the disk
$\sigma_R(0)$.

\citeasnoun{kuij1995} also give a numerical program to calculate the
distribution functions and densities. Their algorithm consists of the
following 4 steps. (i) A test potential function $\Psi^{(0)}(R,z)$ is
given. (ii) The potential is substituted into the density functions
which are calculated from the distribution functions. (iii) Poisson's
equation is solved by using \citeasnoun{pren1970}'s method to determine
a new potential functions. (iv) The procedure 2 and 3 are repeated until
the potential functions have converged.

By using \citeasnoun{kuij1995}'s method, we construct the Galaxy models
which satisfy observational properties. We apply the following
constraints;
\begin{itemize}
\item solar radius : $R_0$ = 8~kpc

\item circular velocity of the disk at the solar radius :
  $V_\mathrm{c}=220$ km/s

\item total surface density within 1.1~kpc of the disk plane :

  $\Sigma_{1.1}(R_0) = 71 \pm 6 M_\odot \mathrm{pc}^{-2}$
  \cite{kuij1991b}

\item contribution of the disk material to $\Sigma_{1.1}$ :

  $\Sigma_\mathrm{d}(R_0) = 48 \pm 9 M_\odot \mathrm{pc}^{-2}$
  \cite{kuij1991b}

\item total Galaxy mass within 50~kpc :

  $M_\mathrm{tot}(<50\mathrm{kpc}) = 5.4^{+0.2}_{-3.6} \times 10^{11}
  M_\odot$ \cite{wilk1999}

\item total Galaxy mass within 170~kpc :

  $M_\mathrm{tot}(<170\mathrm{kpc}) = 1.9^{+3.6}_{-1.7} \times 10^{12}
  M_\odot$ \cite{wilk1999}

\item the halo is spherical : $q=1.0$, i.e.,  $A=0$
  \cite{ibat2001}
\end{itemize}

Moreover, we assume that 
\begin{itemize}
\item disk mass is $5\times 10^{10}M_\odot$,

\item disk scale length is 3.2 -- 3.5~kpc,

\item disk scale height is 200 -- 250~pc,

\item disk edge is 25 -- 28~kpc,

\item bulge mass is 15\% of that of the disk,

\item and bulge size is about 2~kpc.
\end{itemize}

\begin{table}[htbp]
  \caption{Sizes of the components in the Galaxy models, which are
    used in this paper}
  \label{tab:models}
  \renewcommand{\tabcolsep}{1.5pc} 
  \renewcommand{\arraystretch}{1.2} 
  \begin{tabular}{@{}lcccc} \hline
    & \multicolumn{2}{c}{\bf Model L }
    & \multicolumn{2}{c}{\bf Model S } \\ 
    \hline
    & mass ($M_\odot$)& radius &  mass ($M_\odot$) & radius  \\  
    Disk & $5 \times 10^{10}$ & 3.2kpc & $ 5 \times 10^{10}$ & 3.5kpc \\
    ~~~Scale height & & 224pc &  & 245pc  \\
    ~~~Disk edge & & 25.6kpc &     & 28kpc  \\
    Bulge & $0.76\times 10^{10}$ & 2.21kpc & $0.75\times 10^{10}$ &
    2.38kpc \\ 
    Halo & $5.11\times 10^{12}$  & 1539kpc & $8.59 \times 10^{11}$ &
    262kpc \\
    \hline
  \end{tabular}
\end{table}

The halo mass and extent is very important, because in Weinberg's
scenario, the halo plays an important role as a mediator between a
satellite and a disk. Observational estimate of the halo mass has,
however, a huge error, so that we examine two extreme cases: one with
the largest halo (Model L), and the other with the smallest possible
halo (Model S).  Construction of these models is made by KD1995's code,
which is available in their web site, and the input parameters are
listed up in Appendix A. Some characteristic parameters for the two
models are given in Table~\ref{tab:models}. The disk scale height and
the edge radius are similar in both models. The disk surface densities
for the model L and S are $\Sigma_\mathrm{d}(R_0)=$ 45.3 and 45.5
$M_\odot \mathrm{pc}^{-2}$, while the total surface densities including
halo contribution $\Sigma_{1.1}(R_0)=$ 65.0 and 69.8 $M_\odot
\mathrm{pc}^{-2}$, respectively. The halo masses for the model L and S
are $M_\mathrm{tot}(<50\mathrm{kpc})=$ 5.58 and 4.93 $\times
10^{11}M_\odot$, and $M_\mathrm{tot}(<170\mathrm{kpc})=$ 2.1 and 0.9
$\times 10^{12}M_\odot$, respectively. The determination of the
parameters was done by manual searching of the input parameters. Those
are not unique solutions, because the parameter space of the KD1995's
model is larger than the number of the constraints. The model L and S
are the first best fitted distributions. The bulge and halo radii listed
in the table are the tidal radii, where their densities fall to zero.
For model L, such a large tidal radius is necessary to ensure that the
halo profile is nearly isothermal beyond the LMC orbit ($\sim 100$~kpc).

\begin{figure}[t]
  \includegraphics[width=14cm]{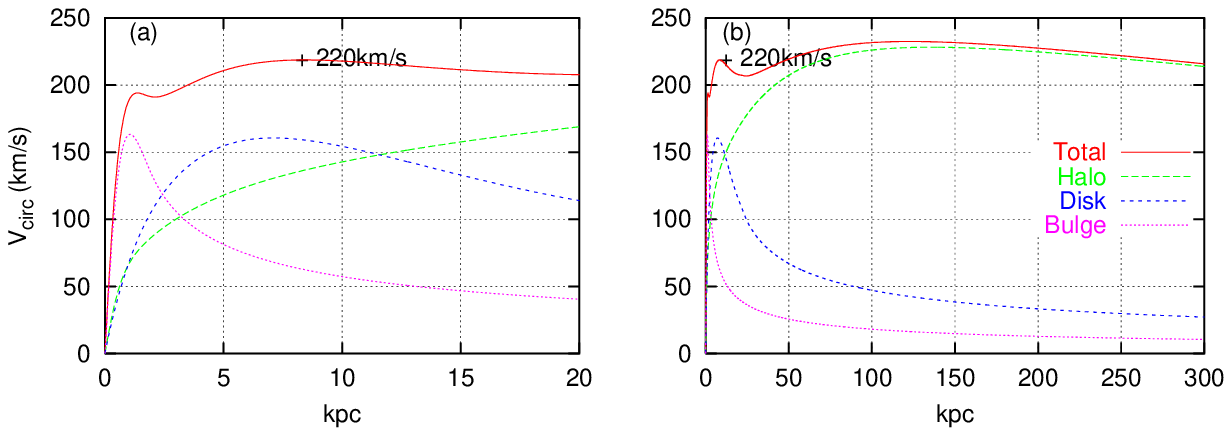}
  \caption{Circular velocity profiles for the model L. Panel (a) shows a
    small scale profile, and (b) shows a larger scale profile. The
    cross sign (+) shows the observational circular velocity at the
    solar radius.}
  \label{fig:modelL_circ_vel}
  \vspace{5mm}

  \includegraphics[width=14cm]{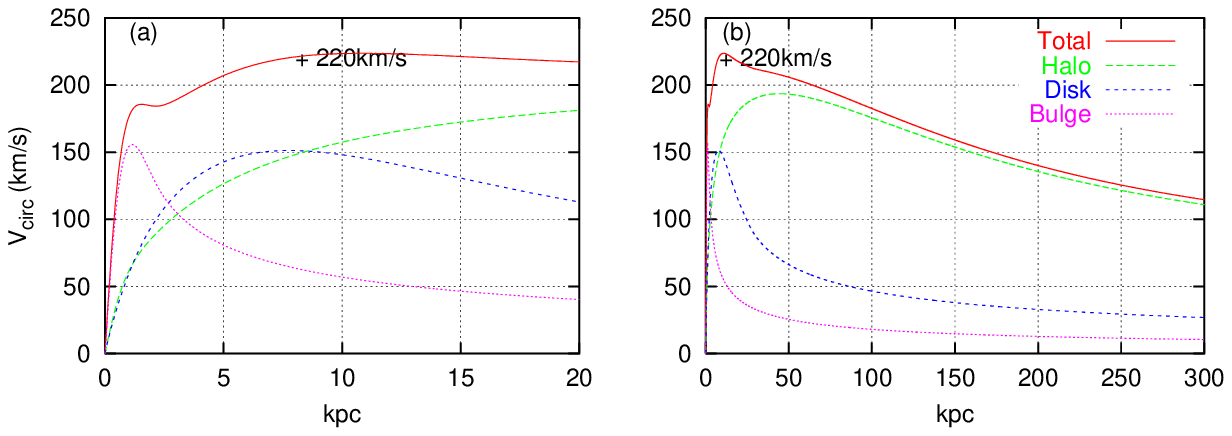}
  \caption{Circular velocity profiles for the model S. Panel (a) shows a
    small scale profile, and (b) shows a larger scale profile. The
    cross sign (+) shows the observational circular velocity at the
    solar radius.}
  \label{fig:modelS_circ_vel}
\end{figure}

\begin{figure}[t]
  \begin{minipage}[t]{60mm}
    \includegraphics{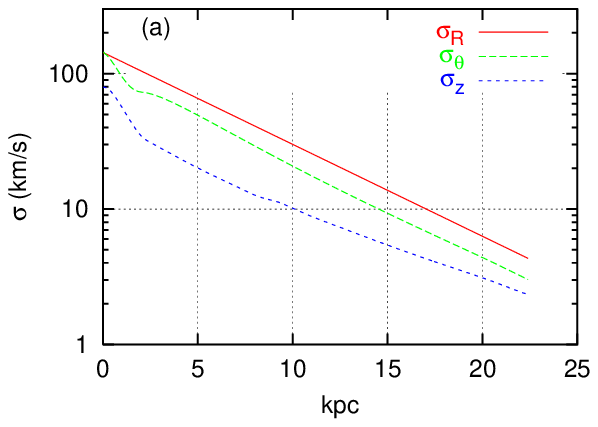}
  \end{minipage}
  \hspace{\fill}
  \begin{minipage}[t]{60mm}
    \includegraphics{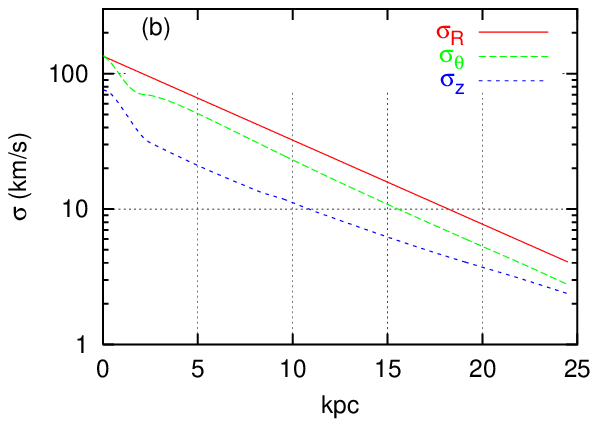}
  \end{minipage}
  \caption{Velocity dispersion profiles of the disk. Panel (a) shows the
    model L and (b) the model S. Three curves correspond to the radial ($R$),
    azimuthal ($\theta$), and vertical ($z$) components.}
  \label{fig:diskvdsp}
\end{figure}

Figures~\ref{fig:modelL_circ_vel} and~\ref{fig:modelS_circ_vel} show the
circular velocity profiles of the Model L and S, respectively.  Within
the disk extent, the circular velocity profiles are almost the same in
the both models, but in outer regions where the LMC is orbiting, the
model L still maintains flat rotation, whereas the model S is nearly
Keplerian. The velocity dispersion profiles of the disk for the model L
and S are shown in Fig.~\ref{fig:diskvdsp} (a) and (b), respectively.
Both profiles are almost the same. The model velocity dispersions at the
solar radius, $(\sigma_R, \sigma_\theta, \sigma_z) \sim (41, 29, 13)$
km/s, are closer to those of the thin disk, and Toomre's Q value has the
minimum (1.7 for the model L, and 1.86 for the model S) around the solar
radius.

Dynamical and numerical stability of the models are shown in
Sec.~\ref{sec:stability}.

We treat the LMC rather simply. It is modeled as an extended single
particle with a spherical Hernquist profile,
\begin{equation}
  \label{eq:hernquist}
  \rho(r)=\frac{M_{\mathrm{LMC}} r_0}{2\pi r(r+r_0)^3},
\end{equation}
where the scale length is chosen to $r_0=3.2$~kpc, then the half mass
radius is 7.7~kpc. The mass of the LMC is set to $M_\mathrm{LMC}=10.6
\times 10^9 M_\odot$, which is twice of the present LMC mass estimated by
\citeasnoun{alve2000}. This large mass includes contribution from the
SMC, which is about 10\% of the LMC, and the mass which is possibly
stripped by tidal field from the Milky Way, during a period of 6 Gyr.

We need to trace the evolution of the Milky Way -- LMC system, at least
6 Gyr to examine the warp dynamics, but it is very difficult to predict
the orbital parameters of the LMC 6 Gyr ago. \citeasnoun{mura1980} have
made a parameter survey with simple assumptions and found that the LMC
has been revolving around the Milky Way in an eccentric orbit
accompanied by the SMC\@. The present position of the LMC is nearly at a
pericenter. With its kinematical data \cite{krou1997a}, the orbital
plane of the LMC is determined. As a realistic assumption of the LMC
orbit, we put the satellite in the present orbital plane at a distance
of 112~kpc for the model L and 97~kpc for the model S. Only tangential
velocity is given so that the initial position will be at apocenter and
the eccentricity is 0.34 and 0.24 for the model L and S, respectively.
As shown in Fig.~\ref{fig:Lsat_radius}, the satellite sinks to below
50~kpc at a pericenter after 6 Gyr. Since our halo is nearly spherical,
the satellite remains in the initial orbital plane.

\section{Numerical Methods}
\label{sec:Numerical_Methods}

\subsection{A hybrid $N$-body code}
\label{sec:Nbody_code}
Numerical simulations for our problem require very large computational
power because of its huge dynamical ranges. One important requirement is to
keep the disk thin enough to distinguish subtle disk vertical dynamics.
Artificial 2-body relaxation is the main source of harmful,
i.e.\ undesired and unphysical heating.
In order to reduce this effect a large number of particles is necessary.
In particular, the relaxation among the disk particles is very large,
because the velocity dispersions are small. Simple application of
Chandrasekhar's relaxation formula \cite{binn1987} reads
\begin{equation}
  \label{eq:relaxation}
  T_\mathrm{relax}=\frac{4.2\times 10^{14}\mathrm{years}}{\ln\Lambda}
  \left(\frac{\sigma}{20\mathrm{km/s}}\right)^3
  \left(\frac{m}{M_\odot}\right)^{-2}
  \left(\frac{n}{\mathrm{pc}^3}\right)^{-1},
\end{equation}
where $\Lambda$ is the Coulomb logarithm, and $\sigma$, $m$, and $n$ are
velocity dispersion, mass, and number density of stars. The critical
particle number in numerical simulations is $10^6$, with which the
relaxation time is $\sim 10^{10}$ years.

For the halo, internal relaxation is no problem, because it is much
hotter than the disk. However, a problem is that the halo is 100 times
more massive than the disk in the model L. Even small noises of the halo
could cause big influence on the disk or the satellite. To reduce the
noise, a large particle number is necessary.

Another difficulty in our simulations is the huge difference in spatial
scales and also in dynamics between the disk and the halo. The most
important feature in the disk is delicate vertical motion, while the
most interesting features in the halo are large scale wakes, which are
excited by the satellite motion. It is of course possible to treat the
system with a single scheme $N$-body code, like a tree-code, but another
possibility is to deal with the disk and the halo by separate
algorithms.

A hybrid $N$-body code, which seems suitable for our system, is the
SCF-TREE code \cite{vine1998}. The disk and bulge are represented by
tree particles (e.g., \citeasnoun{barn1986}), while the halo is treated
by a method expanding the potential using bi-orthogonal polynomial
series, which is widely known as SCF (Self-Consistent Field) method
\cite{clut1973,hernq1992,saha1993}.

A tree code is a rather popular method for general $N$-body problems,
because of its flexibility of system configuration and relatively high
force resolution, which is controlled by softening lengths. The whole
computational space is divided into nested cubic cells to form a
hierarchical tree structure. Interaction between close particles is
calculated by direct summation, but contribution from distant particles
is calculated only as a group of particles contained in a large cell.
This grouping is made with a simple criterion,
\begin{equation}
  \label{eq:opening_angle}
  \theta > \frac{s}{d},
\end{equation}
where $s$ and $d$ are the size and the distance of the grouped cell.
$\theta$ is a control parameter, which is known as the tolerance
parameter. With this grouping the number of calculations reduces to
$\mathcal{O}(N\log N)$ (e.g., \citeasnoun{barn1986}).

A SCF code is, on the other hand, an expansion method for calculation of
gravitational forces. In the code density and potential are represented
by a bi-orthogonal set of basis functions.
\begin{equation}
  \label{eq:expansion}
  \rho(\mathbf{r})= \sum_{nlm}A_{nlm}\rho_{nlm}(\mathbf{r}),  \quad
  \mbox{and} \quad
  \Phi(\mathbf{r}) = \sum_{nlm}A_{nlm}\Phi_{nlm}(\mathbf{r}),
\end{equation}
where each pair $\rho_{nlm}$ and $\Phi_{nlm}$ satisfies Poisson's equation
\begin{equation}
  \label{eq:Poisson_expanded}
  \nabla^2\Phi_{nlm} = 4\pi G \rho_{nlm},
\end{equation}
and bi-orthogonal relations
\begin{equation}
  \label{eq:orthogonality}
  \int \rho_{nlm}(\mathbf{r}) [\Phi_{n'l'm'}(\mathbf{r})]^{\ast}
  d\mathbf{r} =  I_{nl}\delta_{nn'}\delta_{ll'}\delta_{mm'},
\end{equation}
where $I_{nl}$ are normalization factors.

The basic procedure of the SCF method is that first $N$ particles are
distributed to sample the density, then the coefficients $A_{nlm}$ are
calculated by expanding the density distribution. The potential is
determined by the latter equation of (\ref{eq:expansion}), thus
forces can be calculated at any position. The particle positions are
thus advanced to the next time step.

The major advantage of the SCF method is that the speed of the
computation scales as $\mathcal{O}(N)$, which is a smaller burden in
increasing particle number than those of the direct summation method
($\mathcal{O}(N^2$)) or usual tree methods ($\mathcal{O}(N\log N)$). 

A fundamental property of the SCF, which is particular among other
expansion methods, is that each particle contributes not as a local
gravitational source (that is so in other expansion methods), but on
global modes of density distribution, so that there is no
particle-particle interaction in the method. On the other words it is
more suitable for study of the mode interactions. In our problem the
most important dynamics in the halo is dynamical friction on the
satellite.  The classical picture of dynamical friction is due to
scattering of field particles as a 2-body problem
\cite{chan1943,binn1987}. However, modern understanding of dynamical
friction is rather due to interaction between the massive object and
wave modes in the field \cite{trema1984}.  Therefore it seems reasonable
to treat the halo by the SCF method.

In practical computation, the expanded series in eq.
(\ref{eq:expansion}) should be truncated in a small number of terms.
Since computational time is proportional to the number of terms, it is
the most important for the SCF method to find a good basis functions
which can fit the system distribution well with the lowest order term.
There are several basis sets introduced in the literature
\cite{clut1973,hernq1992,saha1993,syer1995,earn1996,robi1996,zhao1996},
but it is still terribly difficult to deal with a complex system like a
disk galaxy.

The SCF-TREE code has strategic advantages more than compromise in our
problem. By using a tree method the disk and bulge enjoy high resolution
and flexibility to possibly large changes in their shape. We do not need
so high resolution in the halo, but large number of particles are
necessary to reduce noise, and the wave modes which are excited by the
satellite motion should be solved correctly. Both requirements are
perfectly accomplished by using the SCF method.

\subsection{Parameter Specification}
\label{sec:Parameter}

Our numerical code is fundamentally the same as
\citeasnoun{vine1998}. The tree part is based on
\citeasnoun{hernq1987}'s algorithm, and for the SCF part
\citeasnoun[hereafter HO]{hernq1992}'s basis set is incorporated. 
There are 7 steps in the calculations at each time-step:
\begin{enumerate}
\item Calculate the coefficients $A_{nlm}$ of the terms in the SCF
  expansion.

\item Build the tree from all the tree particles.

\item Calculate accelerations of TREE and SCF particles by the SCF
  expansion.

\item Form the tree interaction lists among TREE particles, and
  calculate accelerations of TREE particles by the tree system.
  
\item Form the tree interaction lists between SCF particles and TREE
  particles, and calculate accelerations of SCF particles by the tree
  system.

\item Calculate gravitational force from the satellite on all the TREE
  and SCF particles. Acceleration of the satellite is simultaneously
  calculated.

\item Update positions and velocities of all particles and the satellite.
\end{enumerate}

We use a common time step for all particles, and the time integration
scheme is the Leap-Frog. The time steps are fixed to 0.75 Myr and 0.875
Myr for the model L and S, respectively. These time steps are one fourth of
the central free fall time,
$T_\mathrm{ff}\equiv\frac{1}{4}\sqrt{3\pi/(2G\rho_0)} \approx 3$Myr,
which is the shortest time scale in the system. The time resolution at
the center is marginal, but in most part of the disk the time step is an
order of magnitude smaller than the vertical crossing time of the disk.

In the TREE part, $2^{19}(=512\mathrm{k})$ particles are used in total,
and 448k and 64k particles are assigned to the disk and bulge,
respectively. They have the same mass among each component, and mass
ratio between the disk and bulge particles is
$m_\mathrm{bulge}/m_\mathrm{disk}=1.05$. All particles have a single
fixed softening length $r_\varepsilon\approx 50$~pc, which is $1/8$ of
the disk initial thickness. The tolerance parameter is set to
$\theta=0.7$, which is applied to TREE-TREE and also TREE-SCF
interaction lists. In force calculation, up to quadrupole moments are
included.

For the SCF calculations, choice of basis set is the most important. We
employ the HO basis set. Justification of this choice
is shown below.

\begin{figure}[t]
  \includegraphics{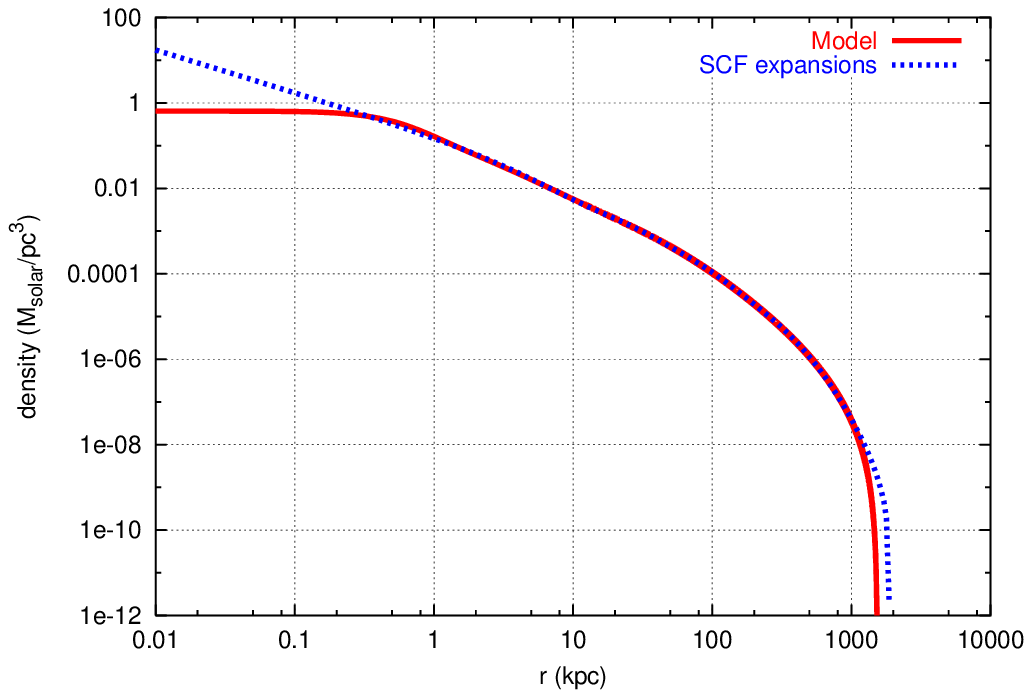}
  \caption{Spherically averaged density profile for the halo of the
    model L, and its SCF expansion with the scale length of the basis
    set $r_0=115$~kpc. Expansion is truncated after $n=8$ term.}
  \label{fig:Lfitting}

  \vspace{5mm}
  \includegraphics{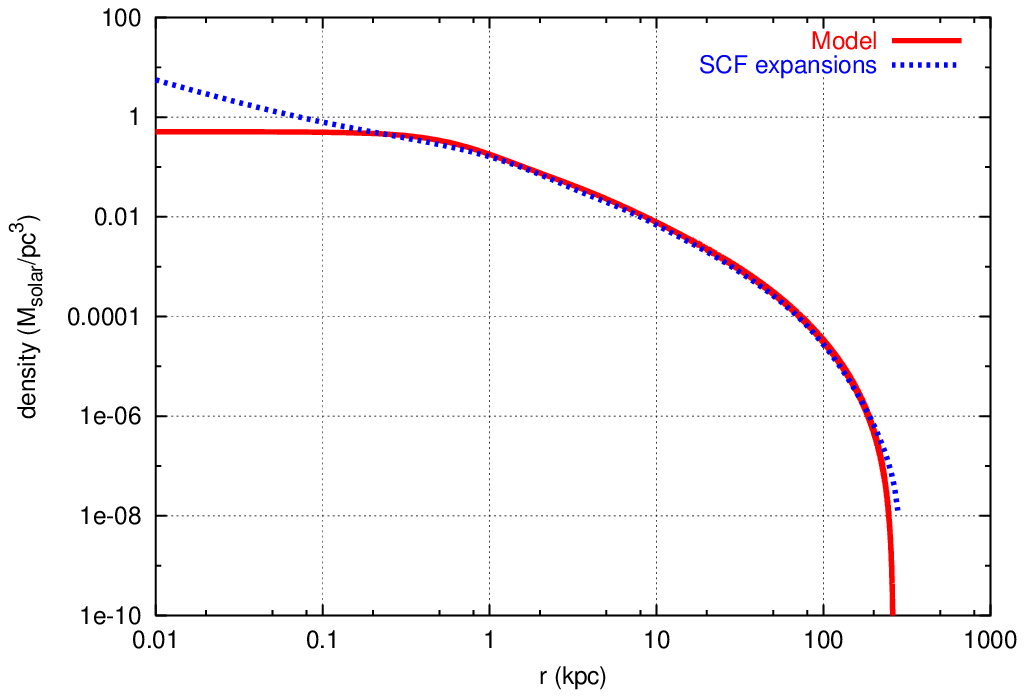}
  \caption{Spherically averaged density profile for the halo of the
    model S, and its SCF expansion with the scale length of the basis
    set $r_0=21$~kpc. Expansion is truncated after $n=9$ term.}
  \label{fig:Sfitting}
\end{figure}

The angular dependence of $\rho_{nlm}$ and $\Phi_{nlm}$ is expanded in
the spherical harmonics,
\begin{eqnarray}
  \label{eq:spherical_harmonics}
  \rho_{nlm}(\vr;\mathbf{r}) &\equiv&
  \sqrt{4\pi}\tilde{\rho}_{nl}(r)Y_{lm}(\theta,\phi) \\
  \Phi_{nlm}(\vr) &\equiv&
  \sqrt{4\pi}\tilde{\Phi}_{nl}(r)Y_{lm}(\theta,\phi),
\end{eqnarray}
where
\begin{equation}
  Y_{lm}(\theta,\phi) = \sqrt{\frac{2l+1}{4\pi}\frac{(l-|m|)!}{(l+|m|)!}}
  P_l^{|m|}(\cos\theta)e^{im\phi}\times 
  \left\{
    \begin{array}{ll}
      (-1)^m & (m\geq0) \\
      1 & ( m<0 )
    \end{array}\right.,
\end{equation}
and the radial functions are expanded by using the Gegenbauer
(Ultraspherical) polynomials,
\begin{eqnarray}
  \tilde{\rho}_{nl}(r) &\equiv& \frac{4n(n+2l+2)+(2l+1)(2l+3)}{4\pi r_0^2}
  \frac{\tilde{r}^l}{(1+\tilde{r}^2)^{l+5/2}} C_n^{(l+1)}(\xi),  \\ 
  \tilde{\Phi}_{nl}(r) &\equiv&  -\frac{\tilde{r}^l}{(1+\tilde{r}^2)^{l+1/2}}
  C_n^{(l+1)}(\xi),
\end{eqnarray}
where
\begin{equation}
  \tilde{r} \equiv \frac{r}{r_0}, \quad \quad
  \xi \equiv \frac{\tilde{r}^2-1}{\tilde{r}^2+1},
\end{equation}
and the convention $G=1$ is adopted. The pair of zeroth order terms in this
expansion has the form
\begin{equation}
  \rho_{000}= \frac{1}{2\pi}\frac{1}{\tilde r(1+\tilde r)^3}, \quad
  \Phi_{000}= -\frac{1}{1+\tilde r},
\end{equation}
which is known as the Hernquist profile \cite{hernq1990a}. This profile
is introduced to fit the light curves of elliptical galaxies. It has a
density cusp at the center, while there is no cusp in our halo model.
This discrepancy causes bad fitting of the basis set in the central
region. Except for that, the expansion with relatively small number of
terms reproduces the original distribution very well.
Figure~\ref{fig:Lfitting} and~\ref{fig:Sfitting} show the results of the
expansion with the number of the radial terms 8 for the model L, and 9
for the model S, respectively. The scale lengths of the basis set are
$r_0=115$~kpc for the model L, and $r_0=21$~kpc for the model S,
respectively.  In both figures the red solid lines show the original
volume density profiles of the models averaged over spherical shells,
and the blue dashed lines show those reproduced by truncated expansion
of the HO basis set. Even with this small number of terms, the
reproduced densities fit the originals fairly well from $r=0.3$~kpc to
nearly the end of the halo extent. The largest discrepancy appears at
the center, where the halo component gives the smallest contribution.
The density of the expanded series becomes comparable to that of the
bulge only within 10~pc from the center. We believe that this difference
has no effect on the warping motion at $R \gtrsim 10$~kpc. In fact we
found no noticeable structural change at the center in our test
simulations owing to the difference of the SCF densities from the
original equilibrium distribution.

Other than the HO basis set, there are several different basis sets
proposed for the purpose of galactic dynamics, including ones with a
finite central core in density profiles. The most well-known example of
the `cored' profile is that based on the Plummer distribution. In this
profile, however, density falls very quickly with increasing radius
($\propto r^{-5}$), therefore the Plummer basis set gives much worse
fitting to our halo profile. In addition, \citeasnoun{zhao1996} gives a
more general class of the basis sets, which include HO and the Plummer
basis sets, but the only profile that has a central core is the Plummer
profile. Therefore the HO basis set seems the best within our knowledge.

Even though the unperturbed halo distribution is approximated well by
small number of the radial functions, the number of terms that is indeed
necessary in our simulations is determined so that the density
perturbations induced by the LMC should be also well approximated. We
have made several test calculations with increasing the radial and
angular functions with the presence of the LMC\@, which is reported in
the next section. We judge the suitable number of terms by examining the
satellite sinking, and select the number of the radial and angular
functions, $n=8$ and $l=12$, respectively. Including more terms does not
affect the satellite sinking history.

The center of the expansion of the SCF basis set is adjusted to the center
of mass of the most tightly bound particles. By using an energy
criterion, about one third of the total halo particles are used to
determine the center of the expansion.

The number of particles that is used to sample the halo density is
$2^{19}$. Therefore we use in total $2^{20}$ particles in the Galaxy in
addition to one extended particle for the satellite. All simulations are
made on Pentium III -- Linux workstations (800MHz). Typical calculation
time is 260 seconds per time step and in total about 500 hours to
complete the evolution up to 6 Gyr. More than 60\% of the CPU time is
used in the TREE part.

\section{Simulations}
\label{sec:Simulations}

\subsection{Stability of the Galaxy models}
\label{sec:stability}
First we examine the stability of our Galaxy models. Without a satellite,
these models must be stable and exhibit no warping motion. Moreover we
have to check the validity of our numerical code. Thus the Galaxy
models are constructed as described in Sec.~\ref{sec:Galaxy_Models} with
in total $2^{20}$ particles, and their evolution is calculated with the
SCF-TREE.

In order to characterize the disk evolution, we make a modal analysis both
of the disk surface density and of the vertical distribution of the
particles. For the disk surface density, $\Sigma(R,\phi)$, angular
dependence is decomposed into the Fourier series,
\begin{equation}
  \Sigma(R,\phi) = \Sigma_0(R) + \sum_{m=1}^\infty \Sigma_m(R)
  \cos(m|\phi-\psi_m(R)|).
\end{equation}
Numerically, the disk particles are binned into annuli, $R \in
[R^{(k-1)}, R^{(k)}]$, where
\begin{equation}
  0\leq R^{(1)} \leq R^{(2)} \leq \ldots \leq R^{(k)} \leq \ldots \leq R^{(K)},
\end{equation}
then averaged coefficients in the annuli are calculated as follows;
\begin{eqnarray}
  \label{eq:sdcoeff}
  \overline{\Sigma}_0^{(k)} &=& \frac{1}{\Delta S^{(k)}}
  \sum_{R^{(k-1)}<R_i<R^{(k)}} m_i, \nonumber \\ 
  \overline{\Sigma}_m^{(k)}\cos(\psi_m)&=&\frac{2}{\Delta S^{(k)}}
  \sum_{R^{(k-1)}<R_i<R^{(k)}} m_i\cos(m\phi_i), \\ 
  \overline{\Sigma}_m^{(k)}\sin(\psi_m)&=&\frac{2}{\Delta S^{(k)}}
  \sum_{R^{(k-1)}<R_i<R^{(k)}} m_i\sin(m\phi_i), \nonumber
\end{eqnarray}
where $\Delta S^{(k)} \equiv \pi \{(R^{(k)})^2-(R^{(k-1)})^2\}$ is the
area of the $k$th annulus, and $m_i$, $R_i$ and $\phi_i$ are the mass, radius
in the disk plane and azimuth of the $i$-th particle.

\begin{figure}[htbp]
  \includegraphics[width=14cm]{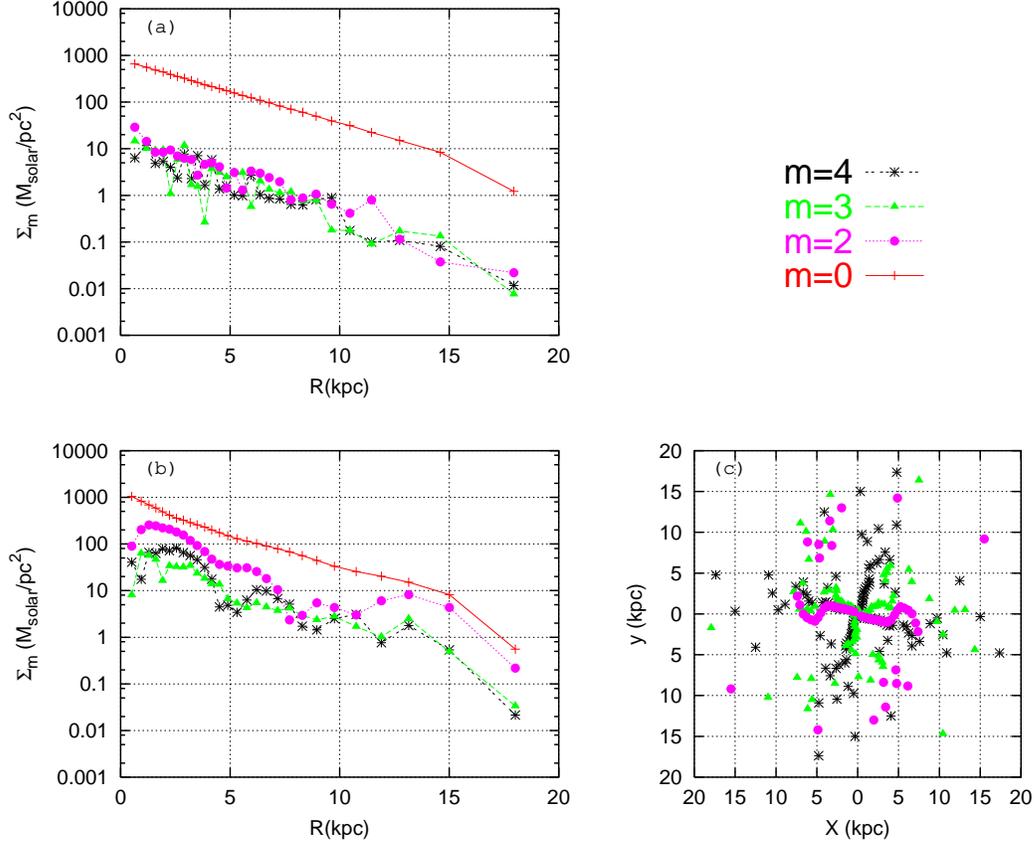}
  \caption{Modal analysis of the disk surface density profile for the
    model L. The panel (a) shows the coefficients of the modes $m=0$,
    2, 3, 4 at the initial time. The panel (b) is the same as (a) but
    after $1.2$ Gyr. The panel (c) shows positions of the peaks of the
    modes $m>1$ at the same epoch as (b). The symbols denote the same
    modes as those in the panel (a). Each annulus contains 16,384
    particles.}
  \label{fig:Lsurfacede0}
\end{figure}

\begin{figure}[htbp]
  \includegraphics[width=14cm]{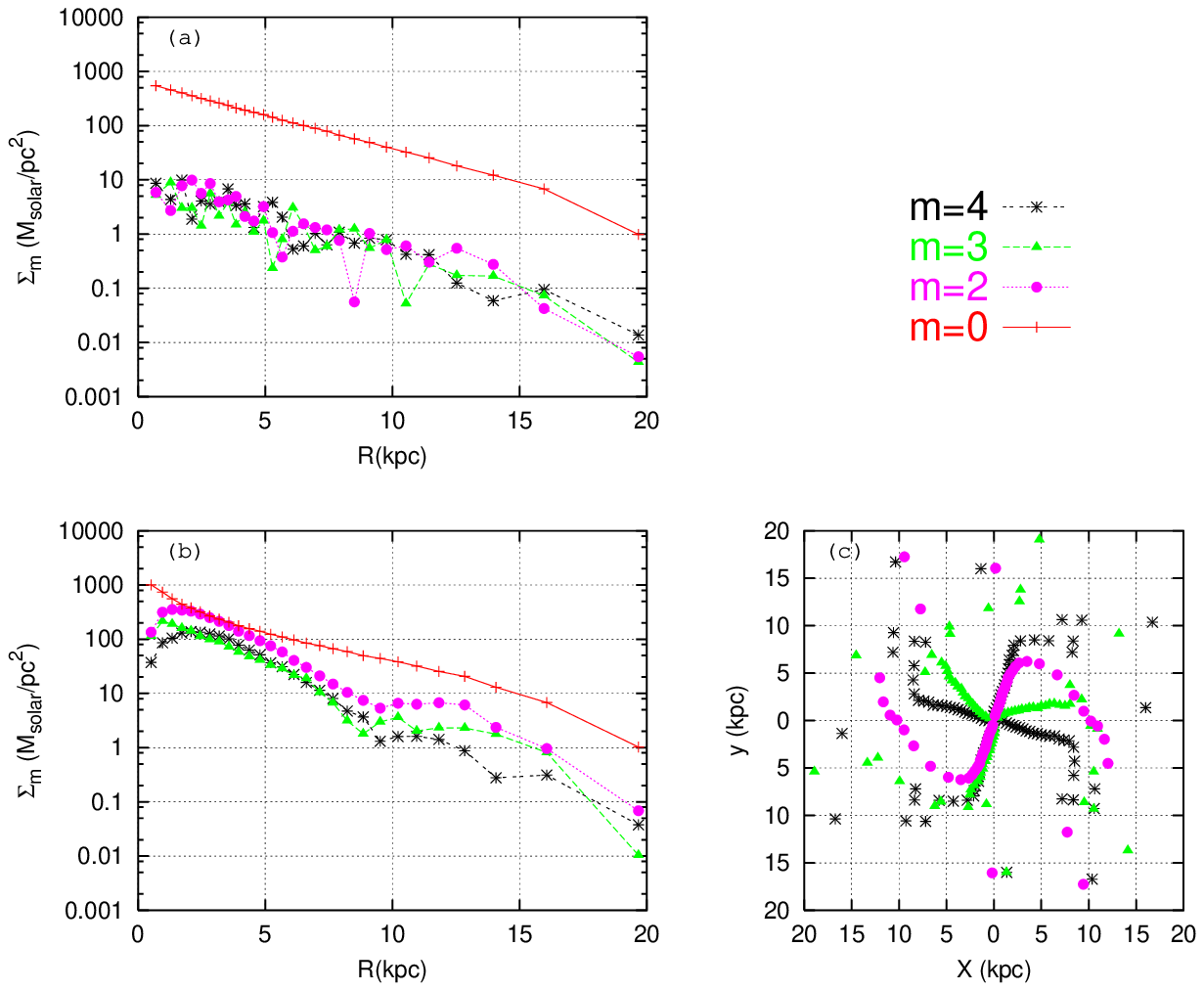}
  \caption{Modal analysis of the disk surface density profile for th
    model S. The notations are the same as Fig.~\ref{fig:Lsurfacede0},
    but the panel (b) and (c) are at $T=1.4$~Gyr.}
  \label{fig:Ssurfacede0}
\end{figure}

Figures~\ref{fig:Lsurfacede0} and~\ref{fig:Ssurfacede0} show the surface
density modes for the model L and S, respectively. The panel
\textbf{(a)} shows the coefficient $\Sigma_m(R)$ at the initial time.
Each annulus contains 16,384 particles. The $m=0$ profiles show the
exponential distribution of the disk surface density.  The higher modes
exist only as noises at the beginning. After about 1 Gyr (panel
\textbf{(b)} and \textbf{(c)}) the profiles show that a bar formed at
the center. The bar in the model S is strong and large, while that in
the model L is 2 or 3 times smaller than in the model S. The formation
of a bar is expected for models with smaller bulges
\cite{ostr1973,sell2001}. We would take the realistic value for the
bulge mass rather than take a larger bulge to prevent the bar instability.
In fact, the Milky Way does have a bar. Therefore our models correctly
reflect the same dynamics as the real Milky Way.

Disk vertical displacement within a annulus $R^{(k-1)}<R<R^{(k)}$ can be
also expanded in a Fourier series;
\begin{equation}
  z_{\mathrm{warp}}^{(k)}(\phi)
  = \sum_m h_m^{(k)} \cos( m|\phi -  \psi_m^{(k)}|)
  = \sum_m \left [ a_m^{(k)} \cos( m\phi )  +  b_m^{(k)}\sin (m\phi) \right].
\end{equation}

A way to determine the coefficients is least squares fitting, in
which the square deviation,
\begin{equation}
  \left( \delta^{(k)}\right ) ^2
    \equiv \langle (z_i - z_{\mathrm{warp}}(\phi_i))^2 \rangle
  \equiv \sum_{R^{(k-1)}<R_i<R^{(k)}} \frac{m_i}{\Delta S^{(k)}
  \overline{\Sigma}^{(k)}(\phi_i)} (z_i -  z_{\mathrm{warp}}(\phi_i))^2 
\end{equation}
becomes the minimum. In this expression, the average, $\langle \dots
\rangle$, is normalized by the local surface density in order to avoid
virtual warping modes owing to the bar or spiral arms. For the sake of
simplicity, we will omit the suffix ${}^{(k)}$ from now on. The conditions
which minimize $\delta^2$ are
\begin{eqnarray}
  \label{eq:least0}
  \frac{ \partial \delta^2}{\partial a_0} &=&
  2 \sum_{m'} \left[  a_{m'} \langle\cos(m'\phi_i)\rangle
    + b_{m'} \langle\sin(m'\phi_i)\rangle \right]
  - 2\langle z_i \rangle =0,  \\
  \label{eq:leasta}
  \frac{ \partial \delta^2}{\partial a_m} &=&
  2\sum_{m'} \left[ a_{m'} \langle \cos(m\phi_i)\cos(m'\phi_i)\rangle
    + b_{m'} \langle \cos(m\phi_i)\sin(m'\phi_i)\rangle \right]
  - 2\langle z_i \cos(m\phi_i)\rangle \nonumber \\
  &=&0, \\
  \label{eq:leastb}
  \frac{ \partial \delta^2}{\partial b_m} &=&
  2\sum_{m'} \left[ a_{m'} \langle \sin(m\phi_i)\cos(m'\phi_i)\rangle
    + b_{m'} \langle \sin(m\phi_i)\sin(m'\phi_i)\rangle \right]
  - 2\langle z_i \sin(m\phi_i)\rangle \nonumber \\
  &=& 0.
\end{eqnarray}
In the analytical limit where $N$-body sampling is fairly well, the
following relations stand
\begin{eqnarray}
  \label{eq:weight1}
  \langle \cos(m'\phi_i)\cos(m\phi_i)\rangle &=& 
  \left\{
    \begin{array}{ll}
      0 \quad & \mbox{for }m'\neq m  \\
      \frac{1}{2} & \mbox{for }m'= m \neq 0 \\
      1 & \mbox{for } m'=m=0,
    \end{array}
  \right.  \\
  \label{eq:weight2}
  \langle \sin(m'\phi_i)\sin(m\phi_i)\rangle &=&
  \left\{
    \begin{array}{ll}
      0 \quad& \mbox{for }m'\neq m  \\
      \frac{1}{2} & \mbox{for }m'= m,
    \end{array}
  \right.  \\
  \label{eq:weight3}
  \langle \sin(m'\phi_i)\cos(m\phi_i)\rangle &=& 0.
\end{eqnarray}

If this is the case, the coefficients are determined by the following equations
\begin{eqnarray}
  \label{eq:coeff0}
  a_0 &=& \langle z_i \rangle, \quad
  a_m = 2\langle z_i  \cos(m\phi_i)\rangle, \quad
  b_m = 2\langle z_i \sin(m\phi_i)\rangle \\
  \delta^2 &=& \left\langle \left(
      z_i - \sum (a_m\cos(m\phi) + b_m\sin(m\phi)) \right)^2
  \right\rangle
  \nonumber \\
  \label{eq:dev}
  &=& \langle z_i^2 \rangle - a_0^2 - \frac{1}{2} \sum \left(a_m^2 +
    b_m^2 \right).
\end{eqnarray}
Here $\delta$ is considered as the disk thickness, which is corrected
by subtracting the effect of warping.

In practical calculations, however, finite $N$-body sampling causes
errors in eqs. (\ref{eq:weight1}) -- (\ref{eq:weight3}), the least mean
square conditions eqs. (\ref{eq:least0}) -- (\ref{eq:leastb}) become
thus a matrix equation. In order to estimate the errors owing to the
coarse sampling, we also find another form of approximate solutions, in
which the higher order coefficients are progressively determined by the
lower order terms.
\begin{equation}
  \label{eq:coeff1}
  a_0 = \frac{\langle z_i \rangle}{\langle 1 \rangle},
\end{equation}
\begin{equation}
  a_1 =
  \frac{\langle z_i \cos(\phi_i)\rangle-a_0\langle \cos(\phi_i) \rangle}
  {\langle \cos^2(\phi_i)\rangle}, \quad
  b_1 =
  \frac{\langle z_i \sin(\phi_i)\rangle-a_0\langle \sin(\phi_i) \rangle}
  {\langle \sin^2(\phi_i)\rangle},
\end{equation}
\begin{equation}
  a_m =
  \frac{\displaystyle \langle z_i \cos(m\phi_i)\rangle-
    \sum_{m'=0}^{m-1} \left [
      a_{m'}\langle \cos(m\phi_i)\cos(m'\phi_i) \rangle +
      b_{m'}\langle \cos(m\phi_i)\sin(m'\phi_i) \rangle \right]}
  {\langle \cos^2(m\phi_i)\rangle},
\end{equation}
\begin{equation}
  \label{eq:coeff2}
  b_m =
  \frac{\langle z_i \sin(m\phi_i)\rangle-
    \displaystyle \sum_{m'=0}^{m-1} \left [
      a_{m'}\langle \sin(m\phi_i)\cos(m'\phi_i) \rangle +
      b_{m'}\langle \sin(m\phi_i)\sin(m'\phi_i) \rangle \right]}
  {\langle \sin^2(m\phi_i)\rangle}.
\end{equation}

Differences between eq. (\ref{eq:coeff0}) and eqs.  (\ref{eq:coeff1})
-- (\ref{eq:coeff2}) are considered as errors.

\begin{figure}[htbp]
\includegraphics{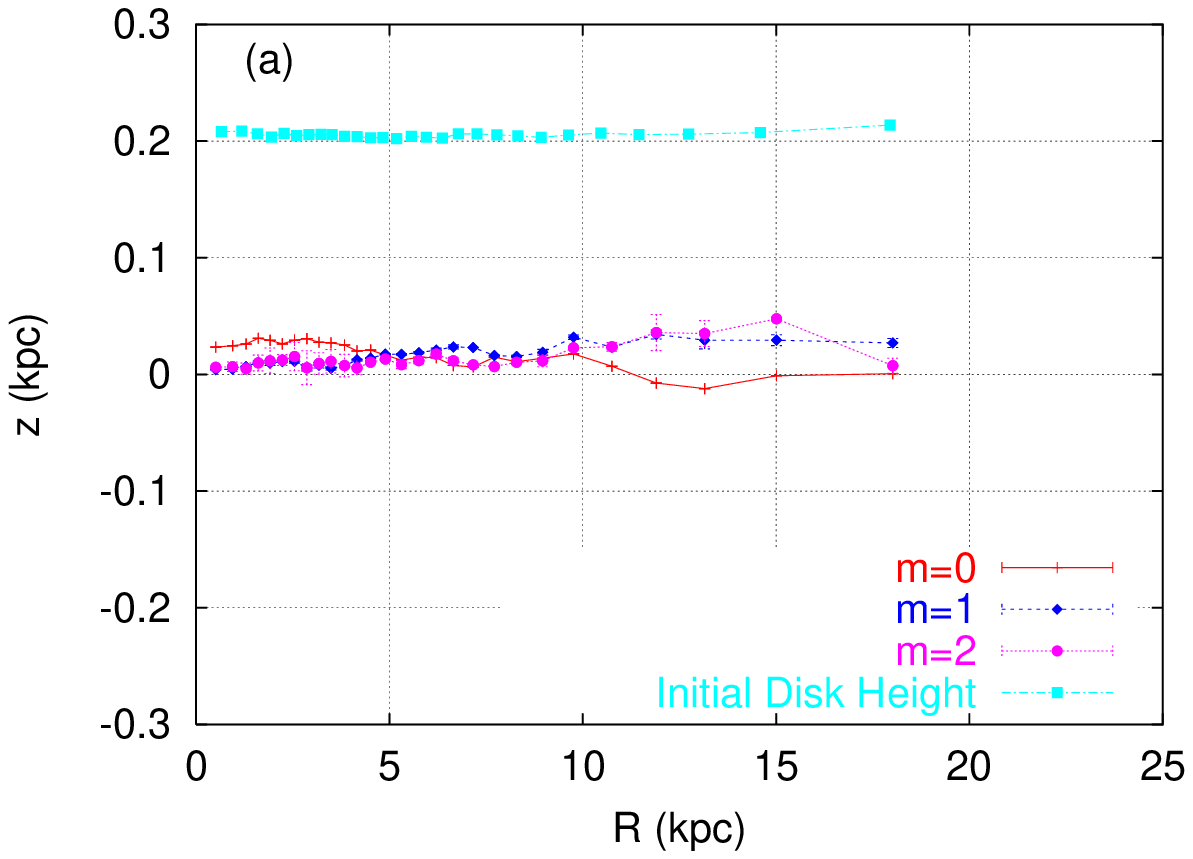}

\vspace{1cm}
\includegraphics{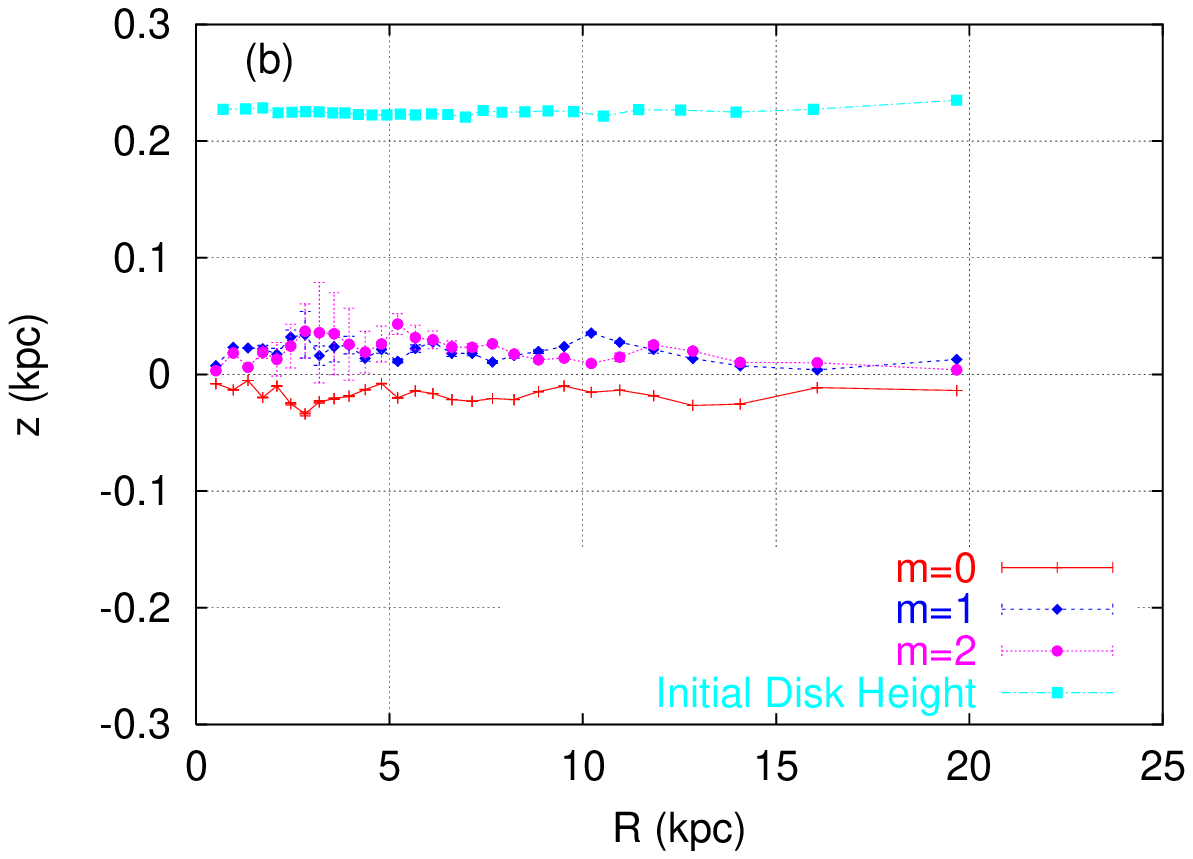}
%
  \caption{Amplitudes of the warping modes of the disk in the model L
    (panel \textbf{a}) and the model S (panel \textbf{b}), at $T=1.2$
    Gyr and 1.4 Gyr, respectively. The modes $m=0$, 1, 2 are shown by
    the curves with plus (+), diamond ($\diamond$), and circle
    ($\circ{} $) symbols, respectively, and the initial disk thickness is
    shown by a curve with box ($\square$) symbols. The error bars denote
    difference between the solutions by eq. (\ref{eq:coeff0}) and eqs.
    (\ref{eq:coeff1}) -- (\ref{eq:coeff2}).
    }
  \label{fig:diskwarp0}
\end{figure}

Figure~\ref{fig:diskwarp0} shows the amplitudes of the warping mode,
$h_m(R)\equiv \sqrt{a_m^2(R)+b_m^2(R)}$, for $m=0$, 1, 2, after about 1
Gyr evolution without a satellite. The errors are shown by the error
bars. Large errors occur at $R\sim3$~kpc in the model S, where the
strong bar appears. In other regions the errors are considerably
smaller, and then invisible in the figures in those scales.

These warping modes are excited only by discreteness noises.
W1998 suggested that the numerical noises could raise a large
warp, if the number of particle in the system is less than 1,000,000.
Comparing with the initial disk heights shown with light blue curves,
it is obvious that the model galaxy disks are kept sufficiently flat over
1 Gyr.

\begin{figure}[t]
  \includegraphics{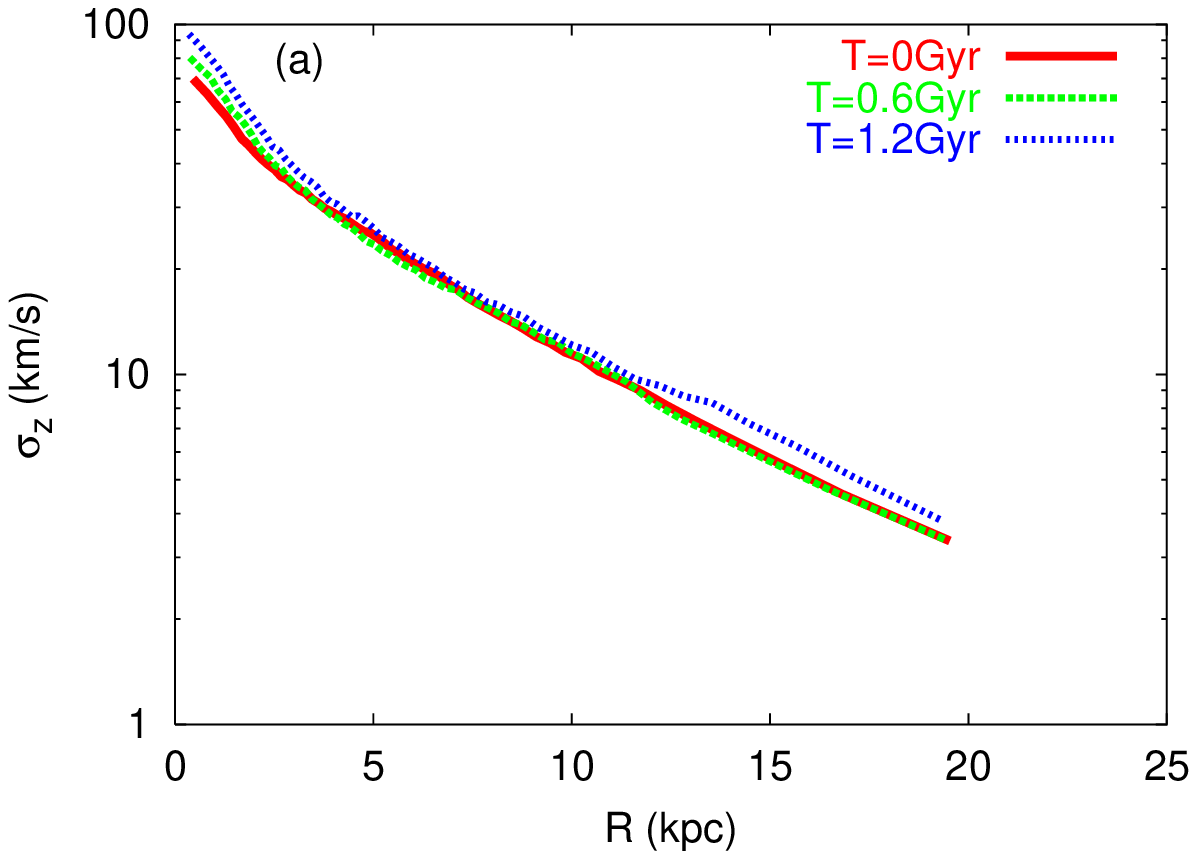}

  \vspace{5mm}
  \includegraphics{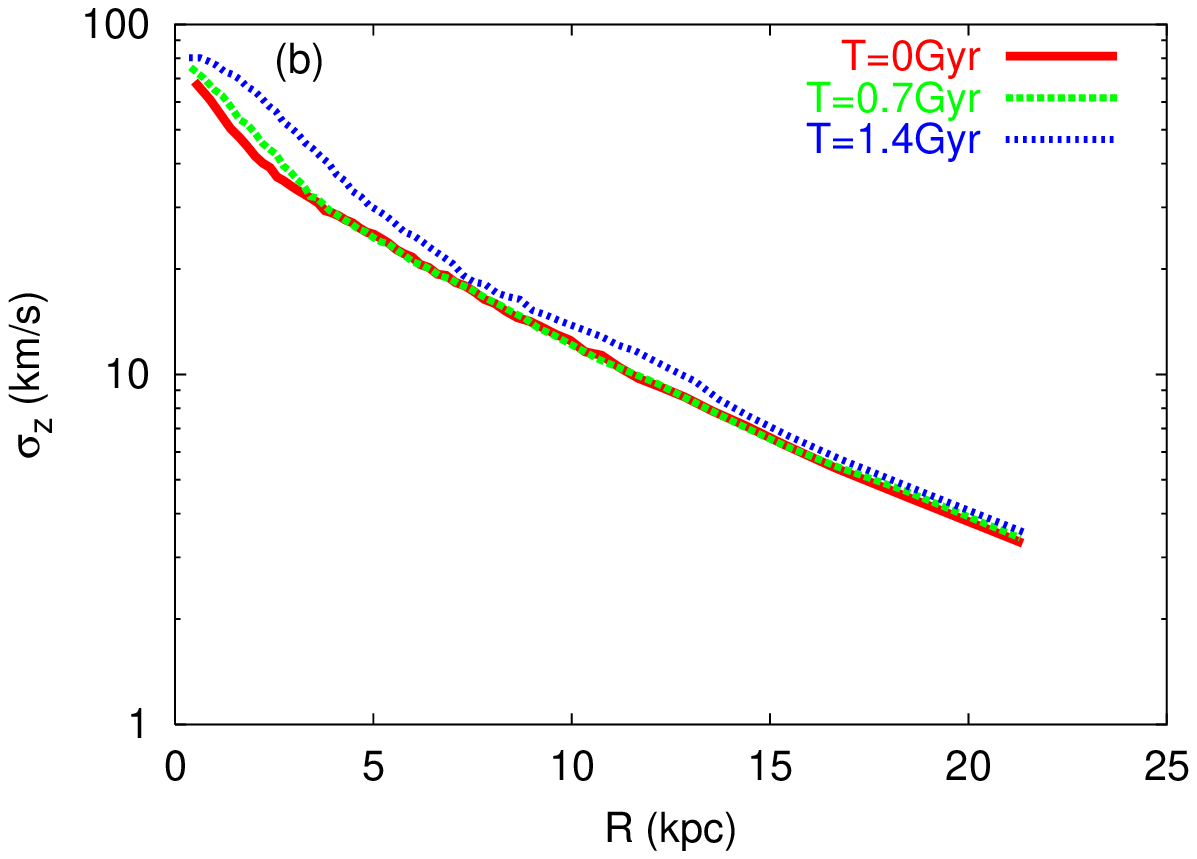}
  \caption{Change in the velocity dispersion in $z$-direction. The panel
    \textbf{(a)} and \textbf{(b)} show vertical velocity dispersion to
    the disk plane ($\sigma_z$) in the model L and S, respectively.}
  \label{fig:veldisp0}
\end{figure}

Two-body relaxation within the disk, and also random noise owing to the
discrete distribution of halo material cause heating of the disk. This
effect is estimated by change in velocity dispersions of the disk
particles. Fig.~\ref{fig:veldisp0} shows the evolution of $\sigma_z(R)$,
which is evaluated in the same annuli as in the warp modal analysis for
the model L and S. In both models, $\sigma_z$ increases significantly in the
central part. This is explained by the disturbance from the bulge.
In particular in the model S, the formation of the strong bar and its
rotating motion might heat up the disk distribution. Except for the
central part ($< 5$~kpc), the increase in $\sigma_z$ is about 10\% (or
1\% in terms of energy increase), after about 1 Gyr evolution. In the
main simulation we will follow the evolution six times longer, but
expected heating would be still harmless.

From these analysis we are convinced that our Galaxy models are stable
enough to discriminate the disk dynamics due to an orbiting satellite.

\subsection{Satellite sinking and dynamical friction}
\label{sec:sinking}

The halo distribution is represented by truncated SCF basis-set. The
necessary number of terms is determined by the requirement that the
response of the halo to the satellite motion should be correctly
solved. In this subsection we examine the nature of the SCF with respect
to the number of terms included in the basis-set expansion.

\begin{figure}[t]
  \includegraphics[width=14cm]{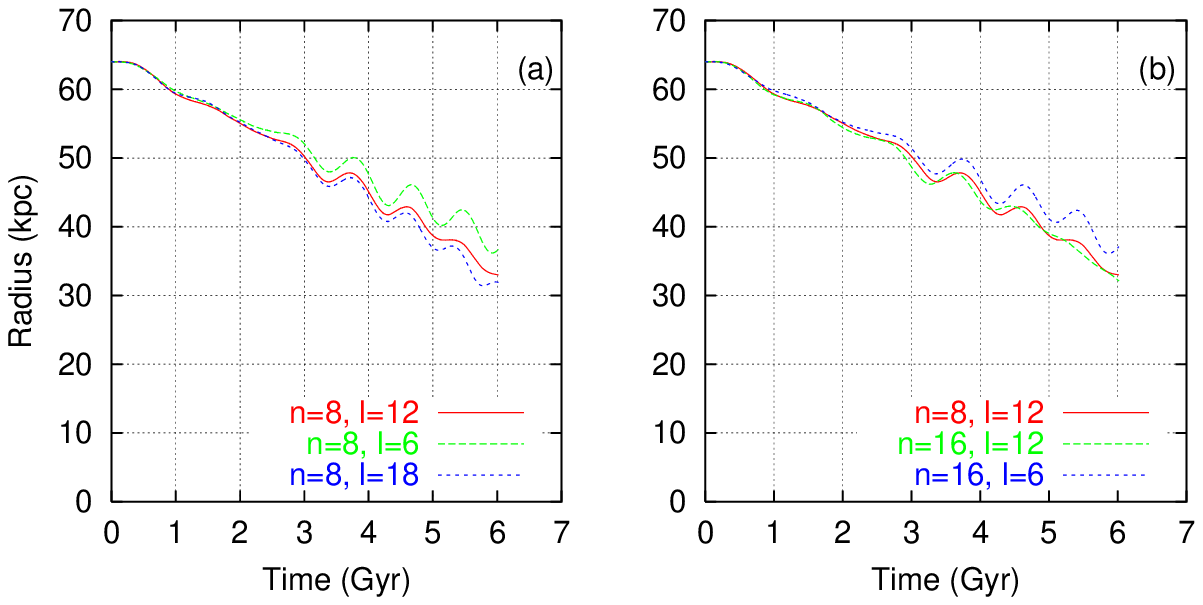}
  \caption{Evolution of the satellite's orbits calculated with different
    SCF parameters. Each parameter of $n$ and $l$ means that the
    terms up to that value of $n$ and $l$ are included in the expansion
    of the basis set.}
  \label{fig:Ldynfric}
\end{figure}

We put the LMC model with a mass of $1.1\times 10^{10}M_\odot$,
initially in a circular orbit with a radius $R=64$~kpc in the symmetry
plane of the model L, but the bulge and disk component are replaced by a fixed
external field. Only the halo component is solved by the SCF method. The
changes in radius according to the satellite sinking are shown in
Fig.~\ref{fig:Ldynfric}.

I have made six different runs with different number of terms in the
expansion. As we have seen in Fig.~\ref{fig:Lfitting}, truncation of the
radial expansion terms at $n=8$ or 9 yields a very good fit to the
unperturbed halo distribution. On the other hand, the dynamical friction
acting on the satellite should be more sensitive to the number of the
azimuthal expansion terms. Fig.~\ref{fig:Ldynfric} (\textbf{a}) shows
the satellite's radius evolution calculated with different $l$ of the
spherical harmonics. All the three runs are made with the radial
expansion terms up to $n=8$, but the spherical harmonics are included up
to $l=6$, 12, and 18. For each $l$, the parameter $m$ runs from $-l$ to
$l$. In the run with $l=6$ the satellite sinking rate is slower
than those with larger $l$. This is because contribution of terms with
higher $l$ to dynamical friction is missing. Increasing $l$ from 12 to
18 makes the sinking rate faster, but the difference is not so
large in comparison with the run with $l=6$. From the figure it is
surmised that dynamical friction with $l=12$ is not far from the true
value. The number of the spherical harmonics up to $l=12$ is the best
compromise between accuracy of the dynamical friction and economy of
the computational time.

In Fig.~\ref{fig:Ldynfric} (\textbf{b}) another comparison regarding the
effect of the radial expansion number $n$ is shown. We find that
including terms larger than $n=8$ does not change dynamical friction so
much. Furthermore, we can make sure that number of the azimuthal
expansion terms is the most important for dynamical friction because the
run with $n=16$ but $l=6$ exhibits still smaller sinking rate.

\subsection{Warp excitation}
\label{sec:warp}

\begin{figure}[t]
  \includegraphics[width=14cm]{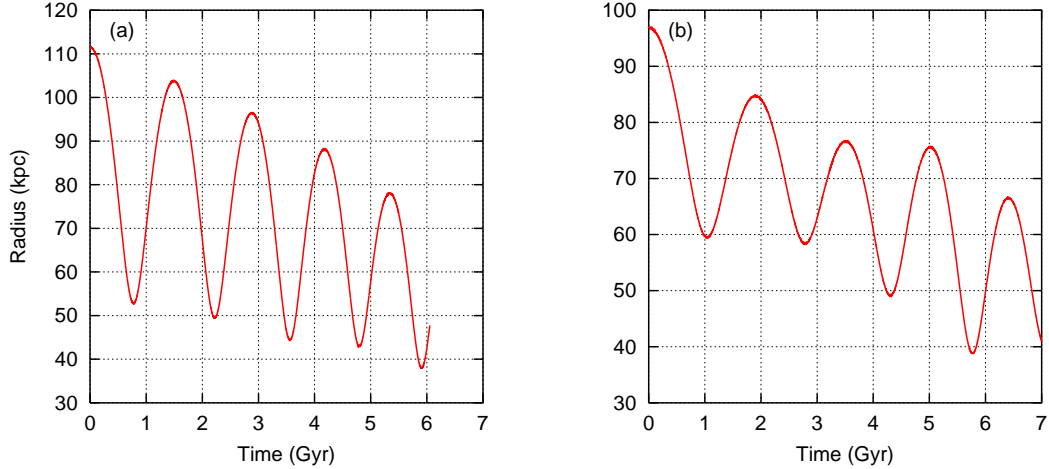} 
  \caption{Change in the radius of the satellite in the model L
    (\textbf{a}) and the model S (\textbf{b}).}
  \label{fig:Lsat_radius}
\end{figure}

Finally we make simulations of the fully live Milky Way models
interacting with the LMC model. Now the satellite is put in the
realistic orbit described in section~\ref{sec:Galaxy_Models}. The
satellite is sinking in the halo owing to dynamical friction. The time
variations in the distance of the LMC from the center of the Milky Way
in the model L and S are shown in Fig.~\ref{fig:Lsat_radius}. In both
models, the LMC is at a pericenter closer than 40~kpc near 6 Gyr.
In fact, the pericenter at 40~kpc is smaller than that of the real
LMC\@. This choice of the orbit makes the gravitational tide a little
bit stronger in our model simulation. We would consider that the moment
of 6 Gyr corresponds to the present day status.  One might notice that
in the model L the LMC sinks regularly, whereas it does not in the model
S. 
Since the halo in the model S has the tidal cut off at 262~kpc, the
satellite is on the edge of the halo distribution.
This is because the halo in the model S has the tidal cut off
around the orbit of the LMC\@. 

\begin{figure}[t]
  \includegraphics[width=14cm]{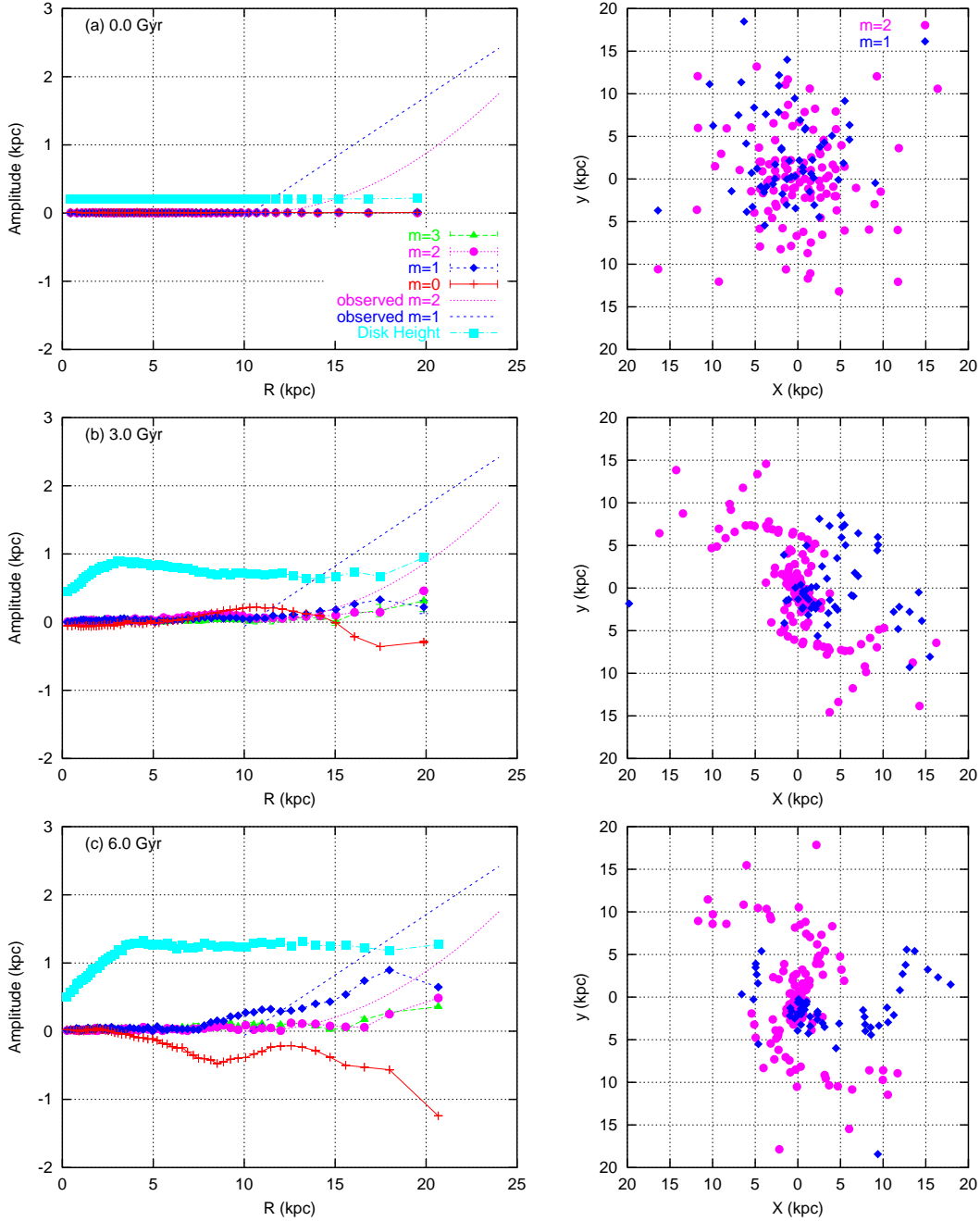}
  \caption{Warp modes in the model L. The rows \textbf{(a)},
    \textbf{(b)}, \textbf{(c)} correspond to the time $T=$ 0, 3.0, and
    6.0 Gyr, respectively. In each row, the left panel shows the
    amplitudes of the warping mode, and the right panel shows the
    position angle of the peak of the warping modes in the disk plane.
    The disk rotation is clockwise in the figures. The LMC orbital plane
    intersects the disk nearly along the $y$-axis in the right panel.
    The amplitude for uniform tilting of the disk is already subtracted
    from $m=1$ amplitudes in these figures.}
  \label{fig:Lwarp}
\end{figure}

Time variation in the disk warping modes in the model L are shown in
Fig.~\ref{fig:Lwarp}. Three rows \textbf{(a)}, \textbf{(b)}, and
\textbf{(c)} correspond to the time $T=$ 0, 3.0, and 6.0 Gyr. In each
row, the left panel shows the amplitudes of the warping mode with $m
\leq3$, together with the disk
thickness. The right panel shows the position angle of the peak of the
warping modes for each annuli in the disk plane. The diamond (blue) and
round (pink) symbols denote $m=1$ and 2 modes, respectively.  The disk
rotation is clockwise in the figures. The LMC orbital plane intersects
the disk nearly along the $y$-axis in the right panel. The first two
intersections of the LMC with the initial disk plane are at ($X$,
$Y$)=(6.8, -91.8)~kpc and (-6.8, 89.3)~kpc. At the beginning (panel
\textbf{a}) all the warp modes are only due to noise, and thus the
position angles are randomly distributed.

\begin{figure}[t]
  \includegraphics[width=14cm]{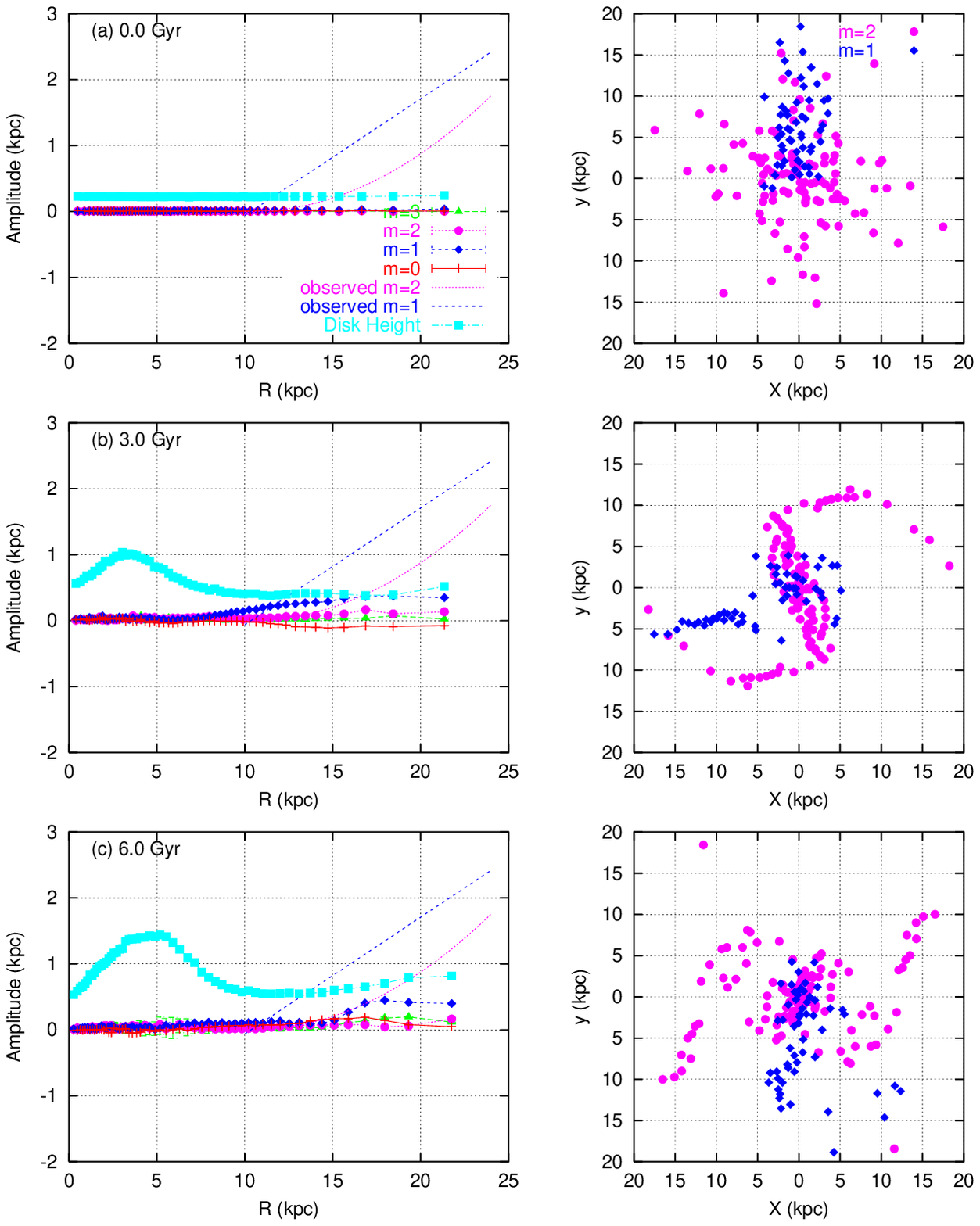}

  \caption{Warp modes in the model S. Notations in this figure are
    the same as Fig.~\ref{fig:Lwarp}. The error bars seen in the left
    panel (\textbf{c}) are the errors in the surface density correction
    on the estimate of warp modes, which is explained in
    section~\ref{sec:stability}. }
  \label{fig:Swarp} 
\end{figure}

The observed warp amplitudes are also shown in the left
panels for comparison. For the Milky Way, not only the $m=1$ but also
$m=0$ and 2 modes are known to exist. We take an approximated formula
for the warp amplitude given by \citeasnoun{binn1998}, with a correction
for the solar radius $R_0=8$~kpc,
\begin{equation}
  z_\mathrm{warp,obs}(R,\phi) = \frac{R/\mathrm{kpc}-10.4}{5.6}\sin\phi +
  0.3\left(\frac{R/\mathrm{kpc}-10.4}{5.6}\right)^2 (1-\cos2\phi).
\end{equation}
In Fig.~\ref{fig:Lwarp} and \ref{fig:Swarp} the observed $m=1$ and 2
modes are shown by dashed and dotted curves, respectively.

At 3 Gyr (panel \textbf{b}) disk warping appears in several modes. The
most noticeable mode is $m=0$ (red solid line with ``$+$'' symbol). It
has a wavy figure with the amplitudes about 250~pc upward and 400~pc
downward. All the other modes have simple shapes with nearly same
amplitudes. The amplitudes grow outwards, but the biggest amplitudes are
still less than 500~pc. The $m=1$ mode shows a clear bending in the
amplitude at around $R=10$~kpc, and the peaks of the $m=1$ mode make a
broad ridge, which slightly winds towards the same direction as the disk
rotation. The $m=2$ global bending mode is clearer in the position angle
diagram which starts at 5~kpc, and winds in trailing sense. But outside
of 11~kpc the mode becomes dispersed again, and shows no coherent
feature.

\begin{figure}[th]
  \includegraphics{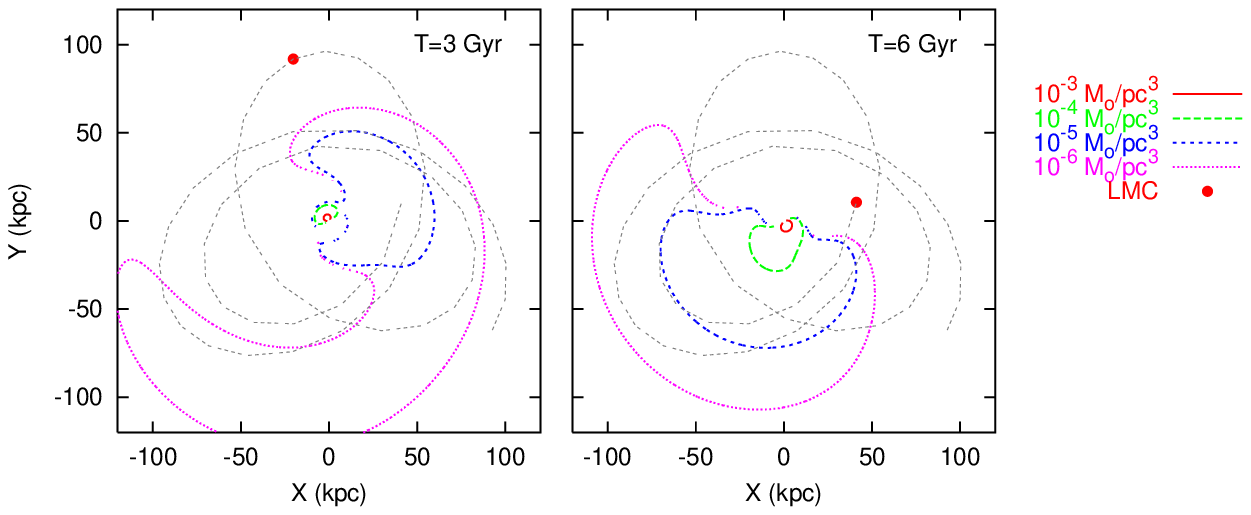} 
  \caption{Density contour of the $m=1$ wake in the model L, in the
    section of the orbital plane of the LMC\@. The position of the LMC is
    denoted by a big dot, where its trajectory is shown by the dashed
    curve. The levels of the contours are shown in logarithmic density
    scales only for positive density enhancement.}
  \label{fig:Lwake1}

  \vspace{1cm}

  \includegraphics{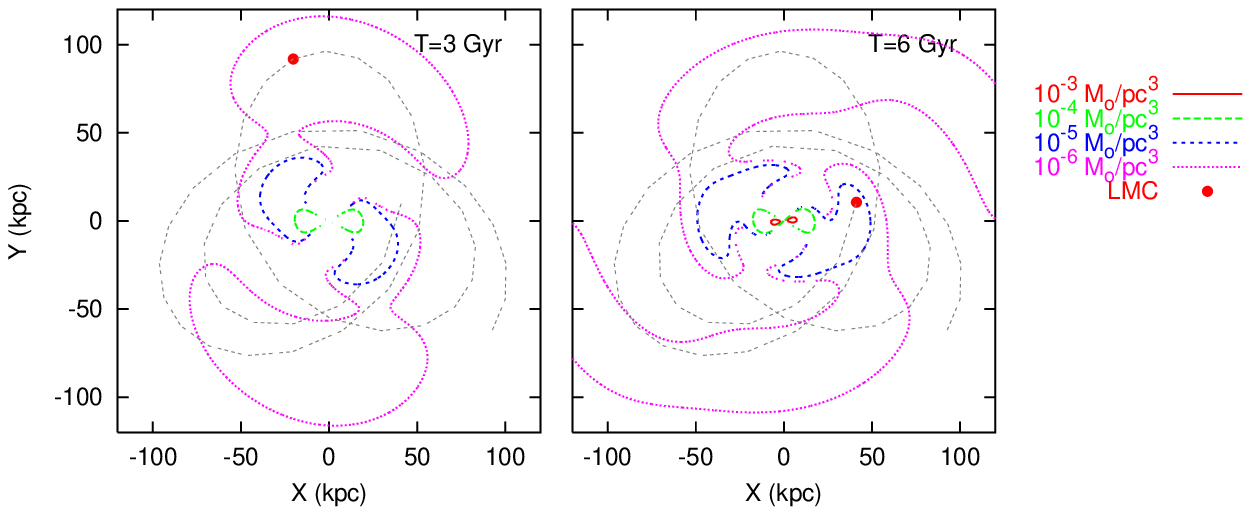} 
  \caption{Density contour of the $m=2$ wake in the model L. Notation is
    the same as Fig.~\ref{fig:Lwake1}. The enhancements that align to
    the horizontal axes in the central region with $R<20$~kpc are
    contribution from the disks.}
  \label{fig:Lwake2}
\end{figure}

\begin{figure}[th]
  \includegraphics{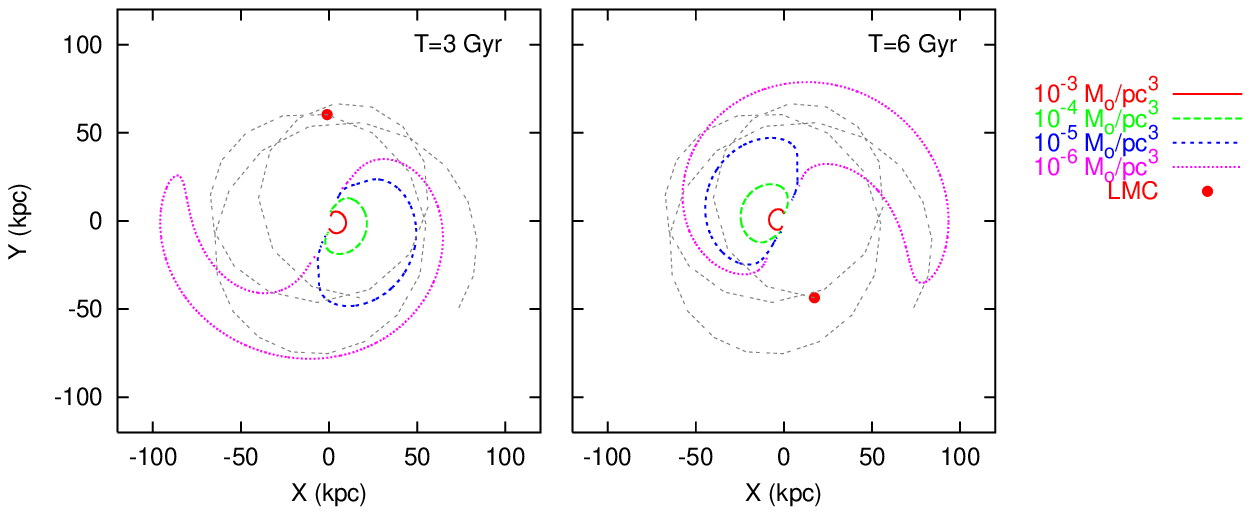} 
  \caption{Density contour of the $m=1$ wake in the model S. Notation is
    the same as Fig.~\ref{fig:Lwake1}.}
  \label{fig:Swake1}

  \vspace{1cm}

  \includegraphics{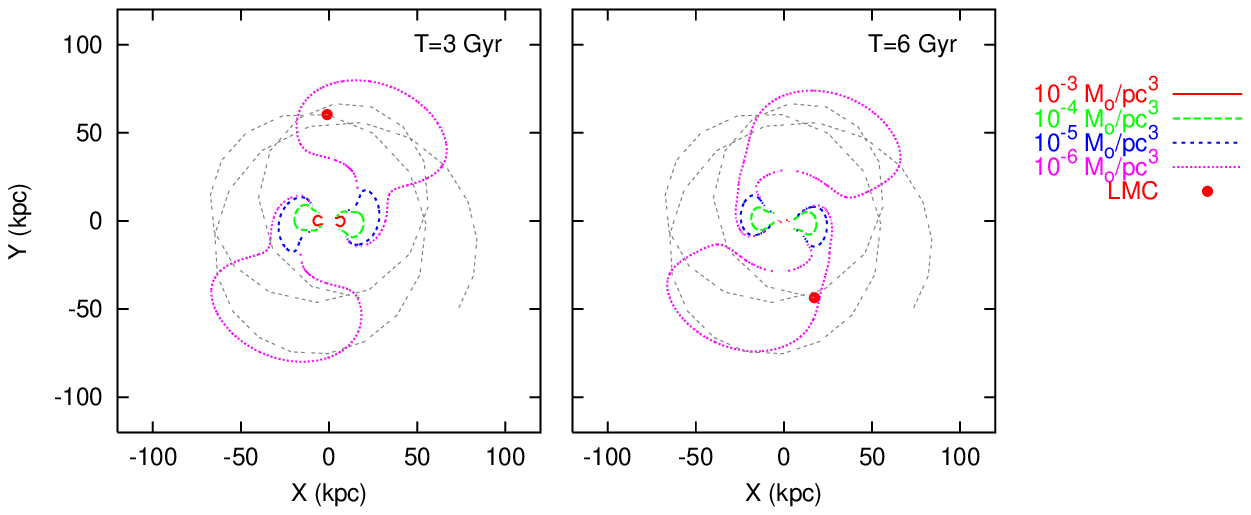} 
  \caption{Density contour of the $m=2$ wake in the model S. Notation is
    the same as Fig.~\ref{fig:Lwake1}. The enhancements that align to
    the horizontal axes in the central region with $R<20$~kpc are
    contribution from the disks.}
  \label{fig:Swake2}
\end{figure}

At 6 Gyr (Panel \textbf{c}), which is considered as the present state,
the amplitudes of the $m=0$ and 1 modes become larger. The largest
amplitude of the $m=0$ is about 1~kpc. A bend in the $m=1$ curve appears
at $R=7.5$~kpc, growing outwards up to the largest amplitude of
1~kpc. The ridge of the $m=1$ mode winds in trailing sense until 15~kpc,
then turns to leading sense. Similar winding feature appears in the
$m=2$ mode, which starts at 10~kpc, winds in trailing sense until
10~kpc, then starts winding oppositely. Since the LMC is revolving
retrograde to the disk rotation, I surmise that warping modes are
affected by the trailing wakes of the LMC (see Fig.~\ref{fig:Lwake1} to
Fig.~\ref{fig:Swake2}), and that is the reason of the change of the
trend in winding of the warping modes.

The most remarkable result is that the warp amplitudes we have got from
the simulation are already comparable to the observed ones.  For the
$m=0$ mode the amplitude is nearly the same amplitude as the observed
one, though the sign is opposite. This negative amplitude in the $m=0$
might depend on the initial position of the LMC, which we took arbitrary
in the present orbital plane. The $m=1$ and $m=2$ modes have more than a
half of the observed amplitudes. The observed trend that the $m=2$ mode
has about a half amplitude of the $m=1$ mode is also realized in the
simulation. This result proves that in cooperation with the halo
response the LMC can create enough tide to excite observable warps.

It should be noted that the disk plane is getting inclined homogeneously
with respect to the initial disk plane. The tilting angles of inner part
of the disk are about 3 degrees at $T=3$~Gyr, and 6 degrees at $T=6$~
Gyr. Since the homogeneous tilting is not observed as a warp, these
tilts are already subtracted from the $m=1$ mode. In fact, tilting
itself can be the cause of warp \cite{spar1988} if a halo is
flattened. Our halo models are, however, spherical, so that tilting
cannot develop the integral-shaped warps.

Fig.~\ref{fig:Swarp} shows the warping modes in the model S\@. The
notations are the same as those in Fig. \ref{fig:Lwarp}. At $T=3$ Gyr
(panel \textbf{b}), a coherent $m=1$ bending mode (dashed curve with
filled diamond symbol) is developed. The bend starts at 8~kpc, and the
ridge of the $m=1$ warp makes a straight line. This feature resembles
the observed warp very well, but its amplitude is at largest 400~pc.
This is significantly smaller than the observed warp. The other
modes are much smaller than the $m=1$ mode. At $T=6$ Gyr (panel
\textbf{c}), all the warping modes still have small amplitudes, thus the
development of warps seems already saturated. In the $m=1$ warping mode
there is a step in the amplitudes at 15~kpc. Within this radius the peak
position angles align in a straight line, but outside of the
radius, the peak position angles wind and folded. This might be
also due to the dominant effect of the wakes in the halo.

One thing that is clear by comparing the results of the two models is
the importance of the halo distribution for excitation of warps. Both
models have the same halo density at the solar radius, but the halo in
the model S has a smaller cut off radius, hence the halo in the model S
has smaller densities in the region where the LMC is orbiting. The wakes
excited in the model S are thus smaller than those in the model L.

Figs.~\ref{fig:Lwake1} to~\ref{fig:Swake2} show contour maps of the
$m=1$ and 2 wakes in the section of the halo cut by the orbital plane of
the LMC\@. In the figures the disks are nearly edge on with the
inclination of 80 degrees, and the section of the disk planes are
aligned to the horizontal axes. The wakes are defined as spherical
harmonic components of the halo density. Only positive density regions
are plotted. The position of the LMC is shown by a large dot, and its
trajectory to 6~Gyr is superposed with a dashed curve. In this plane
the LMC revolves anti-clockwise. 

One should note that the amplitude of the $m=1$ mode has uncertainty
depending on the selection of the origin of expansion. Even for the
spherical distribution, deviation of the origin from the center of the
distribution yield virtual $m=1$ mode. For example, the deviation of the
origin of 625~pc causes a $m=1$ mode with amplitude of
$10^{-5}M_\odot\mathrm{pc}^{-3}$ at $r=50$~kpc, which is comparable to
the one shown in Fig.~\ref{fig:Lwake1}. Our choice of the origin of
expansion is the center of mass of the most tightly bound particles,
which is determined in the same way as the SCF expansion. With this
choice of the origin, the error in the determination of the origin of
expansion cannot be larger than the size of the core ($\sim 500$~pc), so
that the errors in the amplitude of the wake are much smaller
than the amplitude of the wake.

The innermost density enhancements with a size of 20~kpc, which are
seen in the $m=2$ plots (Fig.~\ref{fig:Lwake2} and~\ref{fig:Swake2}),
are contributions from the disk. These figures show that large scale
wakes are excited nearly at the radius of the LMC\@.  It is clear that
the wakes in the model L are larger and stronger than those in the model
S. In particular for the $m=2$ wakes, which would be the direct cause of
the integral-shaped warps, two crescent-shaped wakes are seen in
Fig.~\ref{fig:Lwake2}. One is an outer pair with a radius of 70~kpc,
which typically appears in the $10^{-6}M_\odot/\mathrm{pc}^3$ contours,
and the other is an inner pair with a radius of 30~kpc, which are seen
as the contours of $10^{-5}M_\odot/\mathrm{pc}^3$. Those should
correspond to the wakes which have resonant rotational frequencies 1:1
and 2:1 to that of the LMC\@.  This is exactly the same feature as those
obtained in Weinberg's analysis. In the model S (Fig.~\ref{fig:Swake2}),
however, no clear inner wake can be seen other than the disk
contribution. This might be because the halo density is too small and
the disk gravity is still dominating in the region.

\section{Conclusion and Discussions}
\label{sec:Conclusion}

Our study is meant to be an extension of Weinberg's linear analysis
\cite{wein1998c}, in the way that (i) the simulations are made by a
fully-selfconsistent $N$-body code, (ii) the disk is three-dimensional,
and (iii) the Galaxy models are more realistic. The warp amplitude that
we have got in the model L is nearly the same as that of the largest
warp in W1998. An interesting difference between our results and W1998
is that our smaller halo mass model yields smaller warp amplitudes,
while in W1998 larger halo models yield smaller warps. This is, in fact,
explainable as follows. In W1998, different halo masses produce
different rotation curves, so that larger halo mass models reduce the
efficiency of resonant amplification of warps. In our case the masses of
the disk, bulge and halo within $R=50$~kpc are nearly the same between
the models, hence the rotation curves in smaller scales are also the
same. Therefore the resonant structures are not very different between
the models. On the other hand, the mass within $R=170$~kpc is twice as
massive in the model L as that in the model S. In fact the warp
amplitude in the model L is roughly double of the model S. This linear
dependence shows that the mass of a halo in the region where a satellite
is orbiting has a direct effect on warp amplitudes. As a result we
confirmed W1998's prediction that halos play crucial role in warp
excitation.

Our high resolution simulations upon the Galaxy models with reasonably
realistic parameters make it possible to compare the numerical results
with the observations seriously. Our larger halo model yields very large
warp, which is comparable to the observed warp. The $m=0$ harmonic mode
has nearly the same amplitude as the observed one, while those of $m=1$
and 2 are about a half. These values are in fact remarkably close to
those of the observed warp. Now the LMC has revived as a possible
candidate for the main cause of the Galactic warp. Though there are
still problems, e.g., the lines of nodes are not straight, we have
resolved the most serious problem of the interaction scenario, that the
gravitational tide from the LMC would be too small to excite the
observed warp.

In this paper, we examined the LMC as a main contributer to the Galactic
warp. On the other hand, the Sgr dwarf was posed as another candidate.
Though its current mass is much smaller ($\sim 10^9M_\odot$) than the
LMC, its position is closer (20~kpc), thus the gravitational tide is
comparable to that of the LMC. \citeasnoun{jian2000} proposed a
possibility that the Sgr dwarf had $\sim10^{11}M_\odot$ initially.
Moreover, since the orbital plane of the Sgr dwarf is perpendicular to
that of the LMC, the induce warp would have a line of nodes
perpendicular to that induced by the LMC. Therefore examination of
different satellite parameters is certainly important to find out the
origin of the Galactic warp.

We have also found that our smaller halo model cannot bear the observed
warp. This result shows that even in the interaction scenario the halos
play an important role, so that we could use the warp kinematics as a
probe of the halo structure. Our results might suggest that a massive
halo is necessary to excite the observed Galactic halo if its origin is
the LMC.  In order to restrict halo properties, however, we need to
explore wider model parameters. We could change the contribution of the
halo material within the disk radius smaller to examine the maximal disk
models. While \citeasnoun{mera1998b} argued that a maximal disk model is
excluded, a possibility that the Milky Way has a maximal disk is shown
in several papers\cite{sack1997,palu2000,binn2001b}. Moreover the
strongest warp appeared in the maximal disk model in W1998, which
motivated us to investigate the maximal disk models. Another possible
extension of model parameters is the halo flattening. Our selection of
the spherical halo ($q=1.0$) is based on the analysis of
\citeasnoun{ibat2001}, but other analysis shows that the most probable
flatness of the halo is $c/a=0.8$ \cite{olli2000}. As mentioned in
Sect.~\ref{sec:warp} the disk in the model L is getting inclined to not
negligible extent, so that if the halo is flattened the tilting disks
would get additional torque from the halo, which might make the warp
amplitude larger. Since the shape of the halo is still under big
arguments, for example, studies of galaxy formation by means of
cosmological simulations conclude that the galactic halos are much more
flattened such as $c/a\sim0.5$ \cite{warr1992,dubi1994}. Not only the
shape but also the different velocity distribution in the halo may
affect the warp amplitude, because the wake structure might be
different. We will tackle these problems in detail in subsequent papers.

There is another interesting outcome from our simulations besides the
warp. Quite large disk heating has taken place especially in the model
L. Even though we cannot exclude the artifical heating effect in
numerical schemes, since such a large heating is not observed in the
test simulation without a satellite. We surmise that interaction with a
satellite also causes disk heating. The same heating owing to
interaction with a satellite was studied by \citeasnoun{vela1999}, who
claimed that satellites on prograde orbits causes disk heating more than
those on retrograde orbits. Our simulations results larger heating than
\citeasnoun{vela1999}. This is partly because our initial disks are
about three times thinner than those in \citeasnoun{vela1999}. Since the
disk heating is, as mentioned in Sect.~\ref{sec:Nbody_code}, a very
delicate problem for all numerical methods, we need much more careful
treatment for this problem. This will be our next project.

\ack

I am grateful to Steve Vine for providing me with a copy of his
TREE-SCF code, I would also like to thank, Eliani Ardi, Lia
Athanassoula, Andreas Burkert, Walter Dehnen, John Dubinski, Shunsuke
Hozumi, Andreas Just, and Christian Theis, for many enlightening
discussions. In addition Rainer Spurzem and Burkhard Fuchs gave me
numerous valuable suggestions about the structure and content of this
paper. James Binney, as the referee of this paper, gave critical
suggestions that help me to improve the paper.
This work was supported by Alexander von Humboldt foundation.


\clearpage
\appendix

\section{Input parameters for the Galaxy model construction}
\label{sec:parameters}

For construction of the initial Galaxy models, I have used the software
`GalactICS' \cite{kuij1995}. This software is available on their web
site (\url{http://www.astro.rug.nl/\~{}kuijken/galactics.html}). The
software requires several input parameters. Here is the list of the
input parameters. The parameters explained in Sec.~\ref{sec:Galaxy_Models} are listed in the column 1, and the
corresponding variables used in the codes are listed in the column 2.
The column 3 and 4 are the values of the parameters for the model L and
S, respectively. All the values are in the computational units, where
the gravitational constant, the mass of the disk, and the scale length
of the disk are set to unity.

\begin{table}[thbp]
  \caption{Input parameters given to the software GalactICS.}
  \label{tab:input_params}
  \renewcommand{\arraystretch}{1.2} 
  \begin{tabular}[h]{@{}lccc} \hline
    & & \textbf{model L} & \textbf{model S} \\
    \hline
    \textbf{halo} &&& \\
    cental potential $\Psi_0$ & \texttt{psi00} & $-5.0$ & $-4.0$ \\
    velocity scale $\sigma_0$ &\texttt{v0}   & 0.90 & 1.0 \\
    potential flattening $q$ &      \texttt{q}    & 1.0 & 1.0 \\
    core parameter $R_\mathrm{c}^2/R_\mathrm{K}^2$ ${}^{\ast1}$
    &   \texttt{coreparam}   & 0.1 & 0.1 \\
    characteristic halo radius $R_\mathrm{a}$ ${}^{\ast2}$
    &\texttt{ra}   & 0.7 & 0.8 \\
    \textbf{disk} &&& \\
    disk mass $M_\mathrm{d}$ &        \texttt{rmdisk}   & 1.0 & 1.0 \\
    scale length $R_\mathrm{d}$ &     \texttt{rdisk}   & 1.0 & 1.0 \\
    disk radius $R_\mathrm{out}$ &      \texttt{outdisk} & 7.0 & 7.0 \\
    scale height $z_\mathrm{d}$ &     \texttt{zdisk}   & 0.07 & 0.07 \\
    truncation width $\delta R$ & \texttt{drtrunc} & 0.5 & 0.5 \\
    central radial vel.\ dispersion, $\sigma_R(0)$
                          & \texttt{sigr0} & 0.55 & 0.55 \\
    scale length of $\sigma_R^2$ & \texttt{disksr}& 1.0 & 1.0  \\
    \textbf{bulge} &&&\\
    central density $\rho_\mathrm{b}$ &  \texttt{rho1} & 15.0 & 15.0 \\
    cutoff potential $\Psi_\mathrm{c}$ & \texttt{psiout} & $-4.12$ & $-3.1$ \\
    velocity dispersion $\sigma_\mathrm{b}$ & \texttt{sigbulge} & 0.8 & 0.8 \\
    \hline
    radial step size & \texttt{dr} & 0.01 & 0.01 \\
    number of bins &   \texttt{nr} & 51000 & 10000 \\
    Maximum azimuthal harmonic & \texttt{lmax} & 10 & 10 \\
    \hline
  \end{tabular}

  ($\ast$1) $R_\mathtt{K}$ is the King radius 

  ($\ast$2) Characteristic radius is related to the density scale
  $\rho_1$ by the following relation;
  $ R_\mathrm{a}\equiv [3/2\pi G\rho_1]^{1/2}
    \sigma_0 \mathrm{e}^{\Psi_0/2\sigma_0^2}$

\end{table}

\clearpage

\end{document}